\documentclass[12pt]{article}

\usepackage{amsfonts}
\usepackage{amsmath}
\usepackage{amssymb}
\usepackage{graphicx}
\usepackage{color}

\def\be{\begin{equation}}
\def\ee{\end{equation}}
\def\bea{\begin{eqnarray}}
\def\eea{\end{eqnarray}}
\def\beann{\begin{eqnarray*}}
\def\eeann{\end{eqnarray*}}
\def\nn{\nonumber}

\newcommand\qed{\begin{flushright} $\Box$ \end{flushright}}
\newtheorem{pr}{Proposition}
\newtheorem{de}{Definition}
\newtheorem{theo}{Theorem}

\newtheorem{lem}{Lemma}

\def\circMunit{\bigcirc\!\!\!\!\,\!_{^{\mathbf{1}}}\,}
\def\circMa{\bigcirc\!\!\!\!\,\!_{^{a}}\,}
\def\circMb{\bigcirc\!\!\!\!\,\!_{^{b}}\,}

\def\circMn{\bigcirc\!\!\!\!\,\!_{^{\;\!\!n}}\,}

\def\circMM{\bigcirc\!\!\!\!\,\!\!_{^{M}}\,}

\usepackage{epsfig}

\title{Operationalization of Basic Observables in Mechanics}

\author{Bruno Hartmann\footnote{brunohartmannjr@gmail.com}\\
\textit{Perimeter Institute, Waterloo, ON N2L 2Y5, Canada,}\\
\textit{Humboldt University, D-12489 Berlin, Germany}}

\begin{document}

\maketitle

\textbf{Abstract:} This novel approach to the foundation of the physical theory begins with Hermann von Helmholtz fundamental analysis of basic measurements \cite{Helmholtz - Zaehlen und Messen}. We explain the mathematical formalism from the operationalization of basic observables. According to Leibniz resp. Galilei we define energy, momentum and mass from the elemental comparison ''more capability to work than'' (against same test system) and ''more impact than'' (in a collision) and develop Helmholtz program for direct quantification. From vivid pre-theoretic (principle of inertia, impossibility of perpetuum mobile, relativity principle) and measurement methodical principles we derive the fundamental equations of mechanics.

\section{Operationalization}\label{Kap - KM_Dyn - Operationalization}

The objective is a foundation of physics from the operationalization of its basic observables. The problem is to first determine the physical operations really in a strictly physical way and then formulate a mathematical notation. In another work we have developed basic measurements for relativistic motion \cite{Hartmann-SRT-Kin}. We define the spatiotemporal order ''longer than'' by the practical comparison, whether one object or process covers the other. To express its value also numerically (how many times longer) one invents man-made tools and procedures. Without presupposing arithmetic relations (whose origin and status remain unclear) one can manufacture uniform running light clocks and place them literally one after the other or side by side; and then count, how many building blocks it takes for assembling a regular grid which covers the measurement object. These basic physical operations overlap for different observers. We derive the mathematical relations between the physical quantities (of light clock units), the formal Lorentz transformation. That is a strictly physical approach to physics where initially mathematics must remain outside - and then every step where mathematics is introduced requires an extra justification.

\emph{In absence} of interactions we developed relativistic kinematics from the operationalization of length and duration. Its mathematical formulation is known and given for a circularity free foundation of dynamics. Now we develop Helmholtz program of basic measurements for interactions. A basic measurement requires knowledge of the method of comparison (of a particular attribute of two objects) ''$>_a$'' and of the method of their physical concatenation ''$\ast_a$'' \cite{Helmholtz - Zaehlen und Messen}. We start from the basic observables. According to Heinrich Hertz \cite{Hertz - Einleitung zur Mechanik} introduction to theoretical physics the fundamental role of force must be avoided.\footnote{The concept of ''force'' - as it grew out of Newton's axiomatic system - does not apply the category of causality (what is cause; what is effect) properly in mechanics. The reduction of all phenomena onto forces ties our thinking constantly to arbitrary and unsecured assumptions about individual atoms and molecules (their shape, cohesion, motion etc. is entirely concealed in most cases). These assumptions may have no influence on the final result and the latter may be correct; nonetheless details of those derivations are presumably in large part wrong - according to Hertz - the derivation is an illusory proof. Furthermore force is neither directly measurable nor can it be indirectly determined from Newton's framework alone, without further implicit assumptions \cite{Hartmann-diss}.} He outlines a novel treatment of mechanics based on the notion of energy for directly observable phenomena (independent from Newton's equations of motion and broader). Though Hertz did face the difficulty to specify by which direct experiences we define the presence of energy and determine its quantity - without already anticipating the development of other formalisms for mechanics. We develop Hertz approach and accomplish what he did not succeed.

We define the basic observables energy, momentum, mass from elemental ordering relations \{\ref{Kap - KM_Dyn - Basic observable}\}. According to Leibniz one can compare energy sources directly ''$>_E$'' by their effect against the same test system. For the quantification we develop Helmholtz method, according to which a basic measurement consists in a reconstruction of the measurement object with a material model of concatenated units. Luce, Suppes \cite{Luce Suppes - Theory of Measurement} diagnosed until recently that a major hindrance to understanding basic measurements of energy (and momentum) had been the failure to uncover suitable empirical concatenation operations ''$\ast_E$'' for properties ''capability to work'' and ''impact''. ''By concatenation Grassmann understands any kind of... natural connection, as it may occur in all sorts of coaction of different bodies'' \cite{Hertz - Einleitung zur Mechanik}. We let them coact in a calorimeter \{\ref{Kap - KM_Dyn - Physical Model}\}. We construct the model from pre-theoretic building blocks (by coupling one elementary standard processes of irrelevant internal structure) \{\ref{Kap - KM_Dyn - Reference standards}\} and guided by simple measurement methodical principles \{\ref{Kap - KM_Dyn - Basic observable}\}. Then we change to an \emph{abstract physical perspective} and regard all coacting elements therein solely as carriers of ''capability to work'' and ''impact''. By counting these standardized measurement units we determine the \emph{magnitude} of energy and momentum \{\ref{Kap - KM_Dyn - Quantification}\}. From the geometric layout of the building blocks in those models we derive the primary dynamical \emph{equations} \{\ref{Kap - KM_Dyn - Quantity equations}\}. From the underlying physical operations we develop the mathematical variables and operations (with direct interpretation) and ultimately derive the fundamental equations of mechanics.

\section{Basic observable}\label{Kap - KM_Dyn - Basic observable}

From daily work experience and in play (where valid prognoses about natural processes pay off most) one can \emph{notice} the ''impact'' of decelerating bodies and the associated ''capability to work'' \{\ref{Kap - KM_Dyn - Origin}\}. One develops \emph{pre-theoretic comparison} methods by examples and in words. Physicists fix the \emph{conventions} for observer independent and reproducible procedures.

According to principle of inertia the state of motion is preserved unless some body is effected by an external cause \cite{Euler Anleitung}. For collisions of irrelevant inner structure Galilei defines an elementary \emph{ordering criterion}. Let two generic bodies $\circMa$ and $\circMb$ run into each other with initial velocities $v_a$ resp. $v_b$, collide and stick together.
\begin{de}\label{Def - vortheor Ordnungsrelastion - impulse}
\underline{Momentum} is the striking power of a moving body \cite{Wolff - Geschichte der Impetustheorie}. Object $\circMa_{\:\mathbf{v}_{\!a}} \; >_{\mathbf{P}} \; \circMb_{\:\mathbf{v}_{\!b}}$ has more impact than object $\circMb_{\:\mathbf{v}_{\!b}}$ if in an head-on collision test one body \underline{overruns} the other.
\end{de}
If the bound aggregate $\circMa_{\:\mathbf{v}_a} , \circMb_{\:\mathbf{v}_b} \Rightarrow \circMa\ast\circMb_{\;\mathbf{v}=\mathbf{0}}$ moves neither with object $\circMa_{\:\mathbf{v}_{a}}$ to the right nor with object $\circMb_{\:\mathbf{v}_{b}}$ to the left, then their impact is practically the same.

\begin{de}\label{Def - inertia}\label{Def - vortheor Ordnungsrelastion - inertial mass}
\underline{Inertia} is the - passive - resistance against changes of their motion \cite{Peter '69 - Dissertation}. According to Galilei two bodies $\circMa \; \sim_{m^{\mathrm{(inert)}}} \; \circMb$ have equal inertial mass, if in inelastic head-on collision test $\circMa_{\:\mathbf{v}} , \circMb_{-\mathbf{v}} \Rightarrow \circMa\ast\circMb_{\;\mathbf{0}}$ with same initial velocity no one overruns the other \cite{Weyl - Philosophie der Mathematik und Naturwissenschaft}. One conducts a special case of impulse comparison.
\end{de}

For the development of the concept of energy we follow the review of Schlaudt \cite{Schlaudt}. Mach characterizes the everyday pre-scientific notion ''driving force'': Soon after Galilei one did notice that behind the velocity of an object there is a certain capability to work. Something which allows to overcome force. How to measure this ''something'' was the subject of the ''vis viva'' dispute \cite{Mach - Mechanik in ihrer Entwicklung}. It was initially a vague, pre-theoretic notion. It has the peculiar feature - Schlaudt explains - that it cannot be quantified directly but solely by means of its effect. This is not a mathematical problem but a practical, whose solution entails the mathematical expression for force.

According to Leibniz \emph{equipollence} principle ''il faut avoir recours \`{a} l'equipollence de la cause et de l'effect''. For quantification Leibniz further employs the principle of \emph{congruence}. To measure the cause $\mathcal{S}$ (Ursache) by its effect requires: (i) providing a precise standard action which successively consumes the source $\mathcal{S}$, (ii) the cumulative effect of formal repetitions reproduces the effect of $\mathcal{S}$ and (iii) guarantee that all copies of the standard action are congruent with one another \cite{Schlaudt}. In a practical test ''$\sim_{E}$'' they generate the same effect.

Consider a pair of springs $\mathcal{S}$ and $\tilde{\mathcal{S}}$. When they are compressed (and mechanically locked) the charged springs $\mathcal{S}_E$, $\tilde{\mathcal{S}}_{\tilde{E}}$ become possible energy sources. With their charged state we associate a capability to work. We compare it indirectly by measuring their (kinetic) effect against the same reference system \{\ref{Kap - KM_Dyn - Reference standards}\}: we test whether one spring $\mathcal{S}_E$ can catapult the same test particles with larger velocity than the other, weaker spring $\tilde{\mathcal{S}}_{\tilde{E}}$. Without restricting generality the \emph{charged springs} $\mathcal{S}_E\big|_{\mathbf{v}=\mathbf{0}}$, $\tilde{\mathcal{S}}_{\tilde{E}}\big|_{\mathbf{v}=\mathbf{0}}$ may initially be at rest and after pulling the trigger, after expending the associated capability to work (energy) both \emph{discharged springs} $\mathcal{S}$, $\tilde{\mathcal{S}}$ remain at the state of rest. This allows a clear separation. In return the test particles begin to fly apart. With their motion we associate another form of capability of work (kinetic energy), which they can expend against third parties etc. According to Helmholtz measurement principle: the total capability to work is conserved. In a measurement we consume one specific form entirely (e.g. potential energy until a spring is entirely relaxed; kinetic energy until all projectiles come to rest etc.) in a transformation into other forms (preferably carried by separate elements of the system).
\begin{de}\label{Def - vortheor Ordnungsrelastion - energie}
\underline{Energy} is the capability of a separate source or system to work against an external system $\mathcal{G}$. The kinetic, potential, binding etc. form of energy is associated with exhausting a particular condition of the source (motion, configuration size, chemical bound). According to Leibniz one source $\mathcal{S}_E \; >_{E} \; \tilde{\mathcal{S}}_{\tilde{E}}$ has more potential than another source $\tilde{\mathcal{S}}_{\tilde{E}}$ if the effect of source $\mathcal{S}_E$ on the same test system $\mathcal{G}$ \underline{exceeds} the effect of source $\tilde{\mathcal{S}}_{\tilde{E}}$.
\end{de}
We measure the kinetic energy of a moving body $\circMa_{\:\mathbf{v}_{a}}$ by the number of obstacles it overcomes (until all motion stops). We count how many standard springs can be compressed (repeat congruent standard processes) before the body $\circMa_{\:\mathbf{0}}$ stops. The kinetic energy (of projectiles) transforms into potential energy (of the absorber material). If the latter comes in standard portions, which are congruent with one another, our quantification is complete.

\section{Reference standards}\label{Kap - KM_Dyn - Reference standards}

We define basic observables from direct comparisons. For standardization of experiment and measurement one wants to express their value also numerically (''how many times'' more absorption effect or impetus). We specify a reference process as a sufficiently constant representative of ''capability to work'' and ''impact''. Leibniz presents various candidates including the compression of equivalent springs up to a fixed mark.\footnote{Also D'Alembert utilizes in his \emph{Trait\'{e} de dynamique} congruent actions of a spring. This is a very instructive approach, Schlaudt remarks \cite{Schlaudt}. The action is quantified - not by the depth of compression (in one spring) but instead - by the number of springs which are compressed by a fixed distance. In this way one can \emph{disregard} completely from the \emph{inner dynamics} of the compression process.}

\begin{figure}    
  \begin{center}           
  \includegraphics[height=8.0cm]{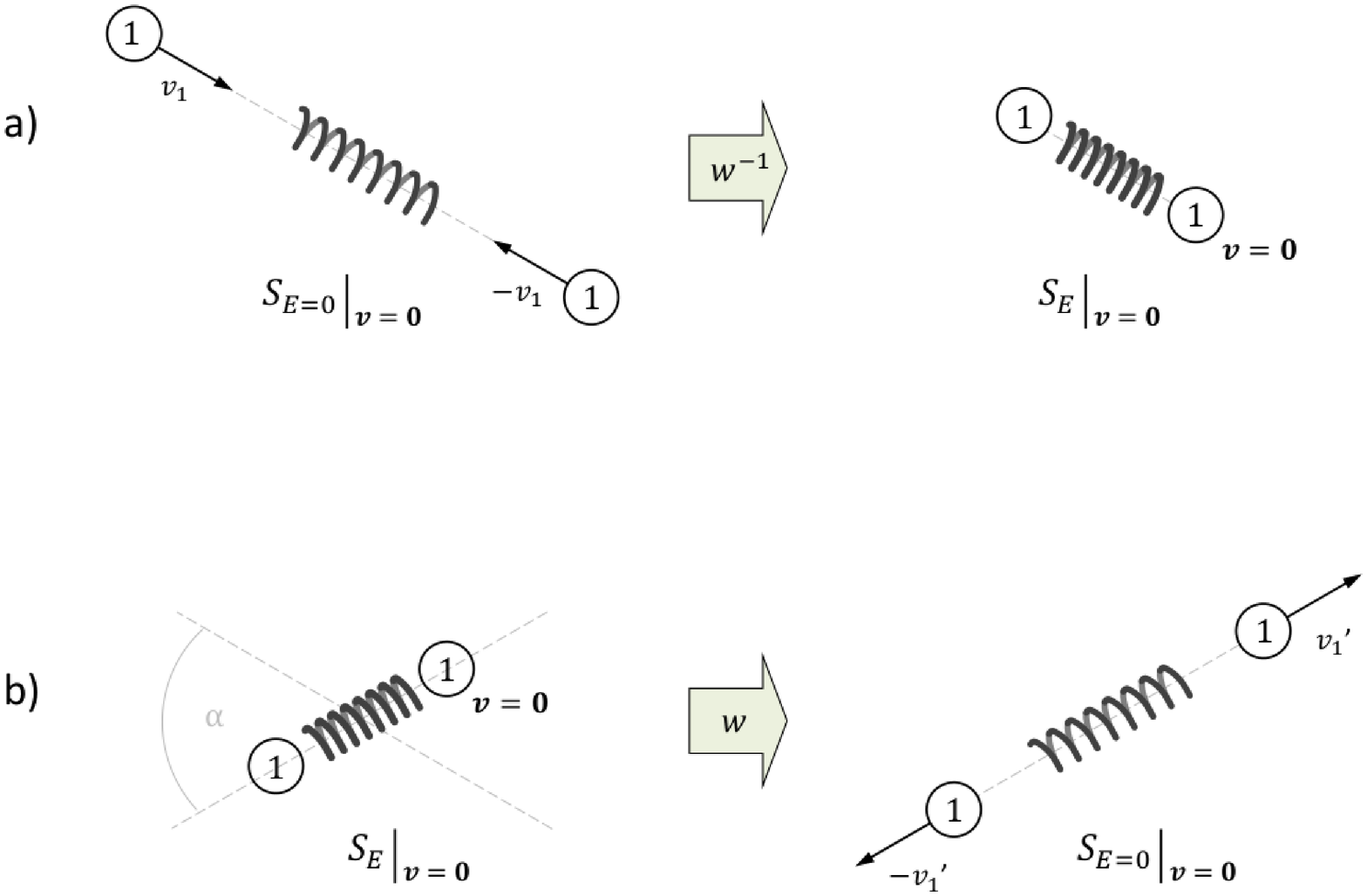}  
  \end{center}
  \vspace{-0cm}
  \caption{\label{pic_Wirkungseinheit_Feder} a) compression, reorientation and b) decompression $w$ of charged spring $\mathcal{S}_E\big|_{\mathbf{0}}$ kicks a particle pair into unit velocity and vice versa (neutral spring $\mathcal{S}_{E=0}\big|_{\mathbf{v}=\mathbf{0}}$ remains at rest)
      }
  \end{figure}

We provide a reservoir $\left\{\circMunit\,, \mathcal{S}\right\}$ with standard bodies ''$\circMunit$'' with same inertia. According to Galilei we can test pre-theoretically if in a head-on collision no one overruns the other. We can charge standard springs ''$\mathcal{S}$'' with same capability to work. Following Leibniz each must catapult our standard objects in the same way. For measurements in entire mechanics we refer to an elementary standard process (of irrelevant internal structure). Let the compressed spring $S_{E=\mathbf{1}}\!\!\mid_{\mathbf{v}=\mathbf{0}}$
\be\label{Abschnitt -- basic dynamical measures - Einheitswirkung}
        \mathcal{S}_{E=\mathbf{1}}\big|_{\mathbf{v}=\mathbf{0}} \,,\: \circMunit_{\:\mathbf{v}=\mathbf{0}} \,,\: \circMunit_{\:\mathbf{v}=\mathbf{0}} \;\;\; \stackrel{w_{\mathbf{1}}}{\Rightarrow} \;\;\; \circMunit_{\:\mathbf{v}_{\mathbf{1}}} \,,\: \circMunit_{-\mathbf{v}_{\mathbf{1}}}
\ee
catapult two resting standard objects into diametrically opposed directions (see figure \ref{pic_Wirkungseinheit_Feder}b)
or reversely let the particle pair (with standard velocity $\mathbf{v}_{\mathbf{1}}$) compress a neutral spring (see figure \ref{pic_Wirkungseinheit_Feder}a). The standard spring $\mathcal{S}_{\mathbf{1}}$ turns standard particles $\circMunit$ into standard impulse carriers $\circMunit_{\:\mathbf{v}_{\mathbf{1}}}$ and vice versa. With regards to the basic observable ''capability to work'' and ''impact'' the inelastic collision $w_{\mathbf{1}}$ is well-defined by symmetry and relativity principle.\footnote{Huygens did study symmetrical collisions between equivalent objects together with the relativity principle to derive the collision laws for Billiard balls. For the same reason Einstein \cite{Einstein '35 - mass energy equivalence} and Feynman \cite{Feynman lectures I} examine interactions between objects which collide and stick together.}

\section{Physical model}\label{Kap - KM_Dyn - Physical Model}

\begin{figure}    
  \begin{center}           
  \includegraphics[height=8cm]{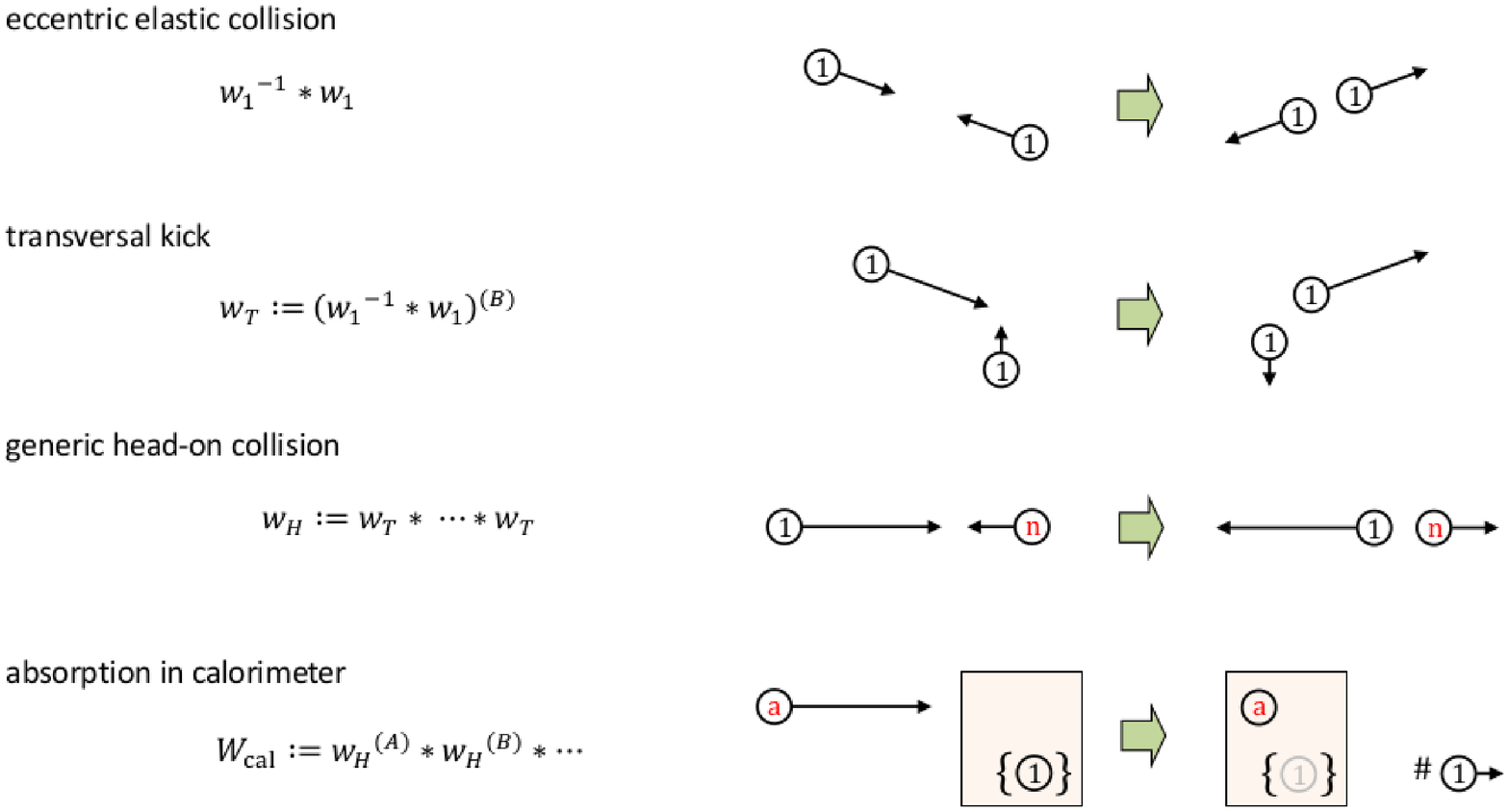}  
  \end{center}
  \vspace{-0cm}
  \caption{\label{pic_Zusammensetzung_Kalorimeter} assemble absorption process $W_{\mathrm{cal}}$ in calorimeter reservoir $\{\circMunit_{\:\mathbf{0}}\}$
    }
  \end{figure}

Consecutive compression and decompression of the spring gives an eccentric elastic collision between bodies of same mass (from the initial state in figure \ref{pic_Wirkungseinheit_Feder}a to the final state in figure \ref{pic_Wirkungseinheit_Feder}b). A drive-by observer will see the process as a transversal kick $w_T$. The kinematics is well-defined by symmetry and covariance. From those standard kicks we assemble increasingly complex collision models (outlined in figure \ref{pic_Zusammensetzung_Kalorimeter}).
By controlled linkage of those building blocks and by relativity principle (view from the moving observer) we model an \emph{elastic head-on collision} $w_H$ between generic (non equivalent!) bodies \{\ref{Kap - KM_Dyn - Collision model}\} and an \emph{absorption process} $W_{\mathrm{cal}}$ for a generic particle in a calorimeter \{\ref{Kap - KM_Dyn - Calorimeter model}\}.

We do not presuppose how velocities of two generic objects change in an elastic collision. The \emph{trick} is to mediate their direct interaction by a steered replacement process with an external reservoir. Our model solely consists of elastic collisions between standard elements which must behave in a symmetrical way. From their layout we derive the generic collision law. In the same way we will derive the mathematical relations for the absorption process in a calorimeter \{\ref{Kap - KM_Dyn - Calorimeter model}\}.

The model building involves a steering task. Physicists couple a series of standard kicks against the respective particles. They concatenate ''$\ast$'' two consecutive collisions $w_{\mathbf{1}}\ast w_{\mathbf{1}}$ in a coinciding element $\circMunit$, which between every two interventions moves freely. A team of assistants steer the initiation timely and at suitable position, such that the desired effect is achieved (see figure \ref{pic_SRT_impulse_reversion}). In every individual action only the final changes in motion matter without needing to know the internal structure of the kicks. Steering the coupling of standard kicks is an elementary operation in a measurement. Like placing the kinematic units, the meter sticks along a \emph{straight} line these practical interventions are pre-theoretic.

\subsection{Elastic collision model}\label{Kap - KM_Dyn - Collision model}

\begin{figure}    
  \begin{center}           
  \includegraphics[height=3.4cm]{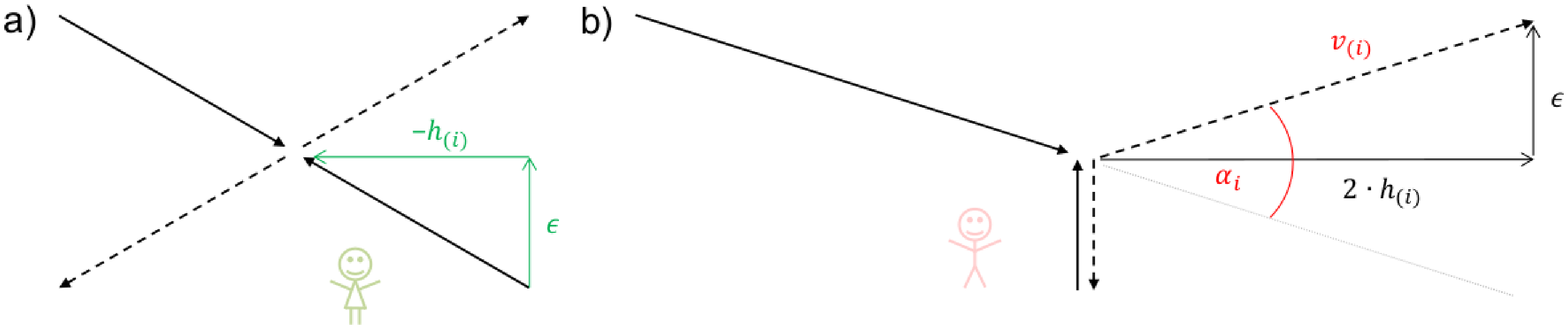}  
  \end{center}
  \vspace{-0cm}
  \caption{\label{pic_elastic_transversal_collision} a) symmetric elastic collision with scattering angle set up by $\mathcal{A}$lice b) appears as elastic transversal collision $w_T$ when $\mathcal{B}$ob drives by with same horizontal velocity to left
    }
  \end{figure}

\begin{lem}\label{Lem - kin quant elast coll - step I Winkel und Geschwindigkeit}
Let in elastic transversal collision between equivalent objects (see figure \ref{pic_elastic_transversal_collision}b)
\be\label{Formel - transversal standard kick}
     \circMunit_{\:\mathbf{v}_{(i)}} \,,\: \circMunit_{\:\epsilon\cdot \mathbf{v}_{\mathbf{1}}} \;\;\stackrel{w_{T}}{\Rightarrow}\;\; \circMunit_{\:\mathbf{v}_{(i)}'} \,,\: \circMunit_{-\epsilon\cdot \mathbf{v}_{\mathbf{1}}}
\ee
reservoir particle $\circMunit_{\:\epsilon\cdot \mathbf{v}_{\mathbf{1}}}$ kick in from below with fixed velocity $\epsilon\cdot \mathbf{v}_{\mathbf{1}}$ and rebound antiparallel. Then incident object $\circMunit_{\:\mathbf{v}_{(i)}}$ moves on with same velocity $v_{(i)}'=v_{(i)}$ deflected by an angle $\alpha_{i}$
\be\label{Formel - v-alpha-elastic transversal collision}
   \sin\left( \frac{\alpha_{i}}{2} \right) = \frac{\epsilon}{v_{(i)}} \;\; .
\ee
\end{lem}
\textbf{Proof:} The elastic collision of identically constituted bodies $\circMunit$ is well-defined by symmetry and Galilei covariance. Let $\mathcal{A}$lice prepare the initial velocities for an eccentric collision\footnote{$\mathcal{A}$lice can freely adjust the deflection $\tan (\frac{\tilde{\alpha}_i}{2}) = \frac{\epsilon}{h_{(i)}}$ by reorienting the spring between two standard processes ${w_{\mathbf{1}}}^{-1}\ast w_{\mathbf{1}}$ (in figure \ref{pic_Wirkungseinheit_Feder}b) or with a suitable impact parameter in an eccentric collision of rigid balls.}
\[
   \mathbf{v}_{(i)} = \left(
               \begin{array}{c}
                 h_{(i)}  \\
                 - \epsilon \\
               \end{array}
             \right) \cdot \mathbf{v}_{\mathbf{1}^{(\mathcal{A})}} \;\;\;\;\;\;\;\;\;\;\;
   \mathbf{v}_{\mathcal{R}} = - \left(
               \begin{array}{c}
                 h_{(i)}  \\
                 - \epsilon \\
               \end{array}
             \right) \cdot \mathbf{v}_{\mathbf{1}^{(\mathcal{A})}}
\]
with fixed horizontal and vertical components (see figure \ref{pic_elastic_transversal_collision}a).

Let $\mathcal{A}$lice move relative to $\mathcal{B}$ob at constant velocity $
   \mathbf{v}_{\mathcal{A}} = \left(
               \begin{array}{c}
                 h_{(i)}  \\
                 0 \\
               \end{array}
             \right) \cdot \mathbf{v}_{\mathbf{1}^{(\mathcal{B})}}
$
in the horizontal direction. Measured values of motion transform by vectorial addition. For $\mathcal{B}$ob incident body $\circMunit _{\: \mathbf{v}_{(i)}}$ has twice the horizontal velocity $2\cdot h_{(i)} \cdot v_{\mathbf{1}^{(\mathcal{B})}}$ with same vertical component $\epsilon\cdot v_{\mathbf{1}^{(\mathcal{B})}}$
\[
   \mathbf{v}_{(i)} = \left(
               \begin{array}{c}
                 2\cdot h_{(i)}  \\
                 \mp\epsilon \\
               \end{array}
             \right) \cdot \mathbf{v}_{\mathbf{1}^{(\mathcal{B})}} \;\;\;\;\;\;\;\;\;\;\;
   \mathbf{v}_{\mathcal{R}} = \left(
               \begin{array}{c}
                 0 \\
                 \pm\epsilon \\
               \end{array}
             \right) \cdot \mathbf{v}_{\mathbf{1}^{(\mathcal{B})}}
\]
while $\mathcal{R}$eservoir particle $\circMunit _{\: \epsilon \cdot \mathbf{v}_{\mathbf{1}}}$ moves up and down vertically with same velocity $\epsilon \cdot v_{\mathbf{1}^{(\mathcal{B})}}$. For the same elastic collision $\mathcal{B}$ob determines a scattering angle $\alpha_{i}$ according to figure \ref{pic_elastic_transversal_collision}b.
\qed
For given initial velocity $v_{(i)}$ and fixed transversal impact velocity $\epsilon\cdot v_{\mathbf{1}}$ we can determine the deflection angle $\alpha_{i}$ - and vice versa provided the latter we find the \emph{admissible velocity} $v_{(i)}$.

\begin{figure}    
  \begin{center}           
  \includegraphics[height=20cm]{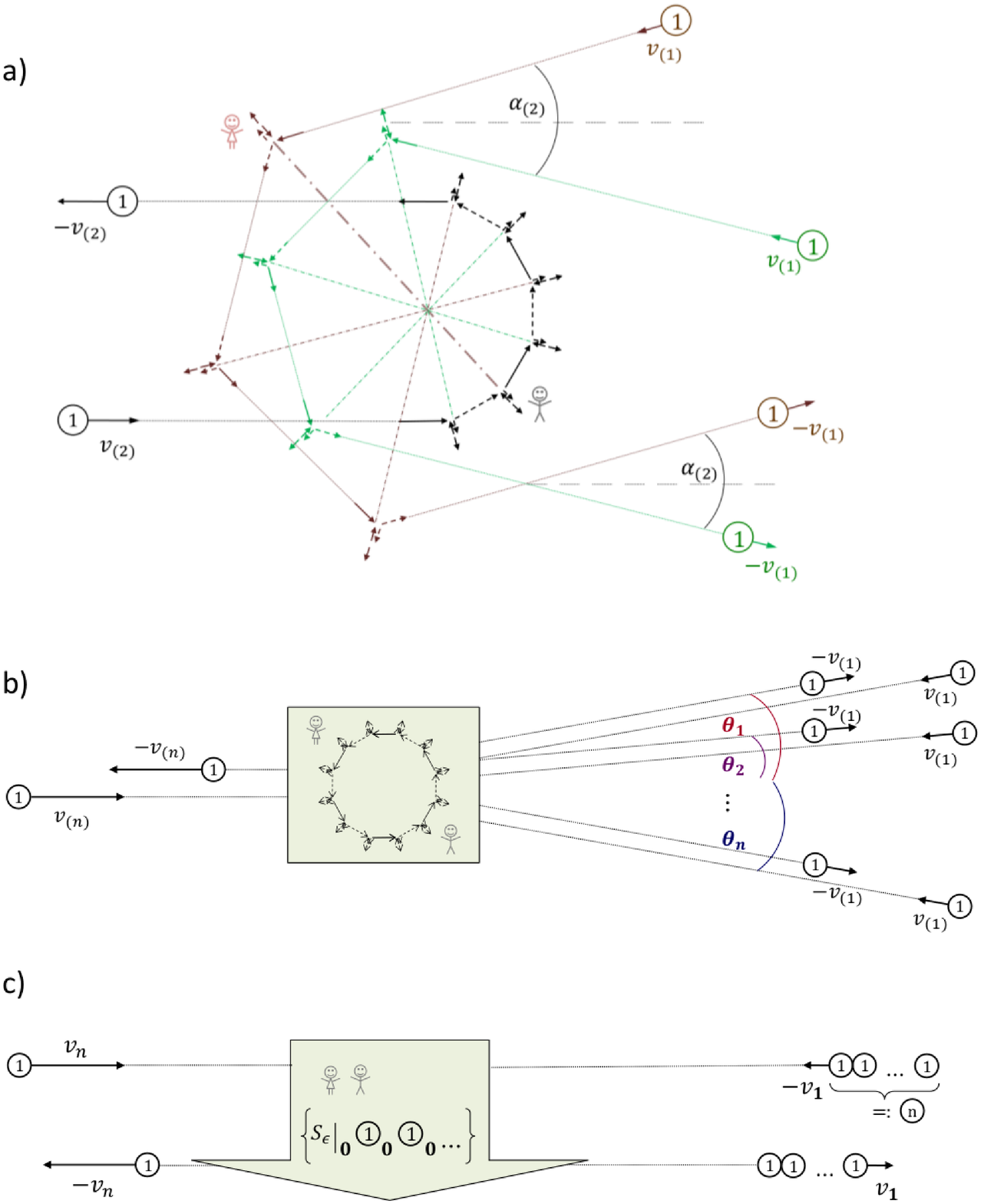}  
  \end{center}
  \vspace{-0cm}
  \caption{\label{pic_SRT_composition_coarse_grained} model of elastic head-on collision between composites of standard objects
    }
  \end{figure}

By a series of transversal standard kicks $w_T$ from reservoir particles we steer a \emph{reversion process} for an incident particle $\circMunit_{\:\mathbf{v}_{(1)}}$ with velocity $\mathbf{v}_{(1)}$ (see figure \ref{pic_SRT_impulse_reversion}); and similar for a faster particle $\circMunit_{\:\mathbf{v}_{(2)}}$ which requires twice the standard reservoir kicks, until its motion is exactly reversed. We align them in the depicted way (see figure \ref{pic_SRT_composition_coarse_grained}a), so that all temporarily mobilized steering elements from the center can be captured again and recycled. In the total balance the reservoir particles do not appear. In the net result \emph{only} the motion of the three incident particles (one from left and two from right side) is exactly reversed. We determine the relation between their \emph{admissible velocities} $v_{(i)}$ from \emph{matching} the building blocks $w_T \left[ \circMunit_{\:\mathbf{v}_{(1)}} \right]$ and $w_T \left[ \circMunit_{\:\mathbf{v}_{(2)}} \right]$ (so that the total configuration functions). By refinement of building blocks we construct similar models for the elastic collision of $n+1$ equivalent particles (see figure \ref{pic_SRT_composition_coarse_grained}b) and in the refinement limit (where the spreading bundle narrows to a ray) for rigid composites of $n+1$ equivalent elements (see figure \ref{pic_SRT_composition_coarse_grained}c).

\begin{theo}\label{Theorem - kin quant elast coll}
Consider a reservoir with identically constituted elements $\left\{\circMunit \right\}$. Suppose we can tightly connect $n$ of them $\underbrace{\circMunit\ast\ldots\ast\circMunit}_{n\times} =: \circMn$ such that the composite acts like one rigid unit. Let in an elastic head-on collision two different composites of standard objects
\be\label{Abschnitt -- kin quant elast coll - elast head-on collision}
   \circMunit_{\:\mathbf{v}} \,,\: \circMn_{\:\mathbf{w}} \;\;\stackrel{w_H}{\Rightarrow}\;\;\circMunit_{-\mathbf{v}} \,,\: \circMn_{-\mathbf{w}}
\ee
repulse from one another with reversed velocities. In Galilei kinematics the velocities satisfy
\be\label{Abschnitt -- kin quant elast coll - kinemtical relations elast collision two generic objects}
   \mathbf{v} \;\; = \;\; - n\cdot \mathbf{w}  \;\; .
\ee
\end{theo}
\textbf{Proof:} We approximate the elastic collision between two generic objects in three auxiliary steps.\footnote{The detailed construction can be thought of as an appendix; the end is marked by the ''$\Box$'' symbol.} Without restricting generality we assume they are composites of unit objects $\circMunit$. We know the collision law for $1+1$ equivalent objects by symmetry and relativity principle (Lemma \ref{Lem - kin quant elast coll - step I Winkel und Geschwindigkeit}). Based on it we construct the collision model for $2+1$ equivalent objects (step I) and for $n+1$ equivalent objects (step II) and ultimately for composites of $n+1$ equivalent objects (step III) to derive the collision law for two generic objects (\ref{Abschnitt -- kin quant elast coll - kinemtical relations elast collision two generic objects}).

In \textbf{step I} we examine the elastic head-on collision between one unit object $\circMunit_{\:v_{(2)}}$ from left with initial velocity $v_{(2)}$ and two unit objects $\circMunit_{\:\mathrm{R}_{15^{\circ}} v_{(1)}}$ and $\circMunit_{\:\mathrm{R}_{-15^{\circ}}v_{(1)}}$ from right with velocity $v_{(1)}$ under suitable orientation $15^{\circ}$ resp. $-15^{\circ}$ (see figure \ref{pic_SRT_composition_coarse_grained}a). We model the process by a series of transversal standard kicks and derive the admissible velocities.

\begin{figure}    
  \begin{center}           
  \includegraphics[height=18.5cm]{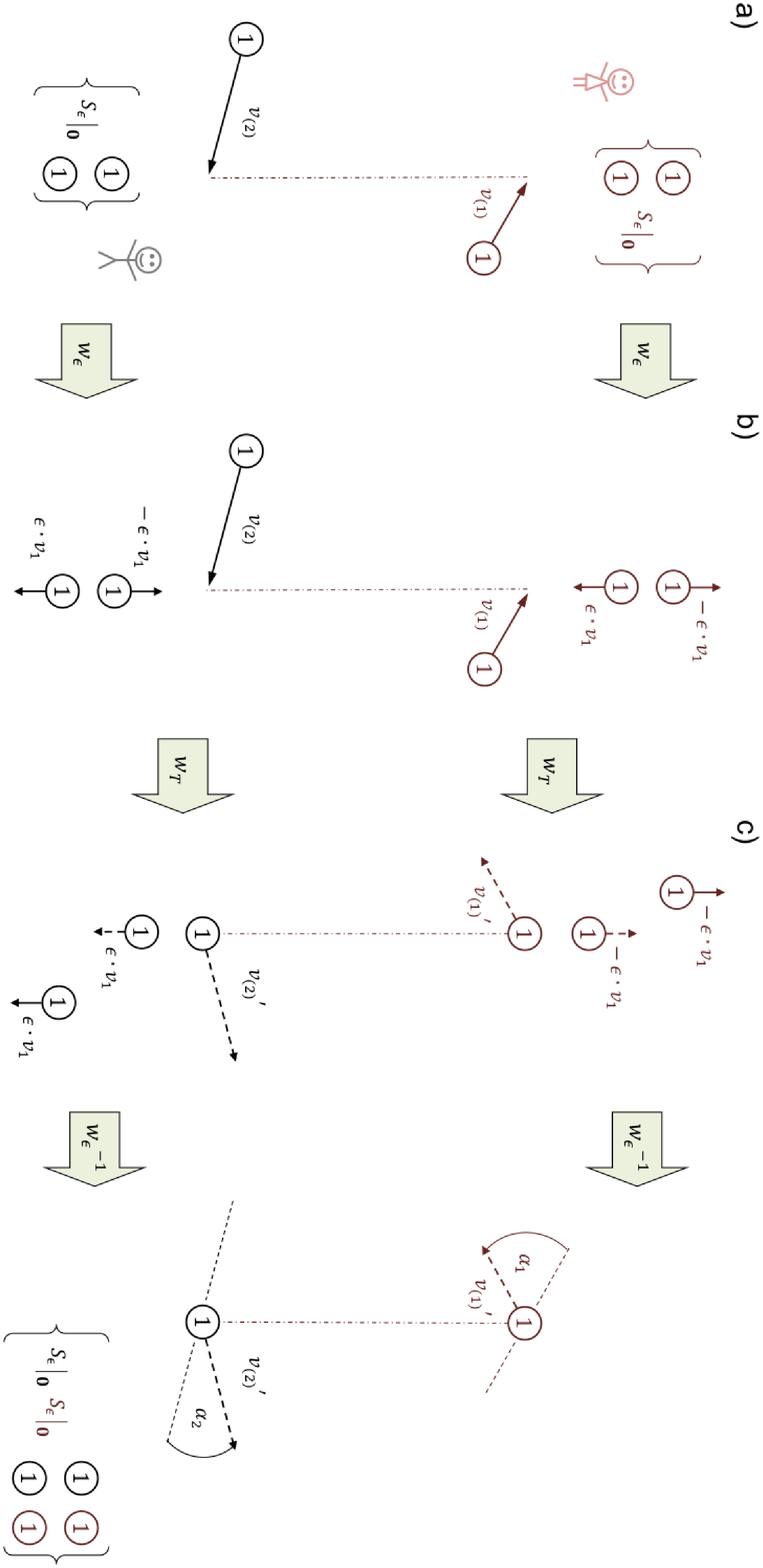}  
  \end{center}
  \vspace{-0.5cm}
  \caption{\label{pic_exact_annihilation1} at diametrically opposed positions Alice and Bob steer a) standard springs $\mathcal{S}_{\epsilon}\:\big|_{\mathbf{0}}$ against resting reservoir elements $\circMunit_{\:\mathbf{0}}$ and $\circMunit_{\:\mathbf{0}}$ to generate a pair of antiparallel impulse carriers for an b) elastic transversal collision with the incident particles $\circMunit\!\!\!\;\!\longrightarrow \!\!\!\!\!\!\!\!\!\!_{v_{\!(1)}}\;\;\;$ resp. $\circMunit\!\!\!\;\!\longrightarrow \!\!\!\!\!\!\!\!\!\!_{v_{\!(2)}}\;\;\;$ c) the antiparallel recoil particles recharge the two temporarily expended springs $\mathcal{S}_{\epsilon}\big|_{\mathbf{0}}$ and return as resting particles back into the reservoir $\left\{\circMunit_{\:\mathbf{0}} \right\}$
    }
  \end{figure}

Let Alice and Bob share an external reservoir $\left\{ \mathcal{S}_{\epsilon}\big|_{\mathbf{v}=\mathbf{0}} , \circMunit_{\:\mathbf{v}=\mathbf{0}} \right\}$. They temporarily expend standard energy sources $\mathcal{S}_{\epsilon}\big|_{\mathbf{v}=\mathbf{0}}$ (of strength $\epsilon$) against initially resting reservoir elements
\be\label{Abschnitt -- kin quant elast coll - prepare congruent transversal impulse}
   \mathcal{S}_{\epsilon}\big|_{\mathbf{0}} \,,\: \circMunit_{\:\mathbf{0}} \,,\: \circMunit_{\:\mathbf{0}} \;\;\stackrel{w_{\epsilon}}{\Rightarrow}\;\; \circMunit_{\,\epsilon\cdot\mathbf{v}_{\mathbf{1}}} \,,\: \circMunit_{-\epsilon\cdot\mathbf{v}_{\mathbf{1}}}
\ee
to \emph{prepare} transversal impulse carriers $\circMunit_{\:\epsilon\cdot \mathbf{v}_{\mathbf{1}}}$ with velocity $\epsilon\cdot v_{\mathbf{1}}$  into suitable direction $\theta$ (see figure \ref{pic_exact_annihilation1}a). They fire them into the momentary way of incident particle $\circMunit_{\:v_{(1)}}$ resp. $\circMunit_{\:v_{(2)}}$ such that the former repulse antiparallel. Each transversal kick $w_T$ successively deflects incident particle $\circMunit_{\:v_{(i)}}$ $i=1,2$ by corresponding angle $\alpha_{i}$ (see figure \ref{pic_exact_annihilation1}b). For \emph{fixed} impact velocity $\epsilon\cdot v_{\mathbf{1}}$ of the reservoir element $\circMunit_{\:\epsilon\cdot v_{\mathbf{1}}}$ and \emph{matching} deflection $\alpha_{1}=60^{\circ}$ and $\alpha_{2}=30^{\circ}$ the admissible velocities are given by $v_{(i)}\stackrel{(\ref{Formel - v-alpha-elastic transversal collision})}{:=}\sin^{-1}\!\left( \frac{\alpha_{i}}{2} \right) \cdot \epsilon\cdot v_{\mathbf{1}}$.

Alice steers the reversion process for incident object $\circMunit_{\:v_{(1)}}$ from the left (see figure \ref{pic_SRT_impulse_reversion}):
\begin{figure}    
  \begin{center}           
  \includegraphics[height=9.2cm]{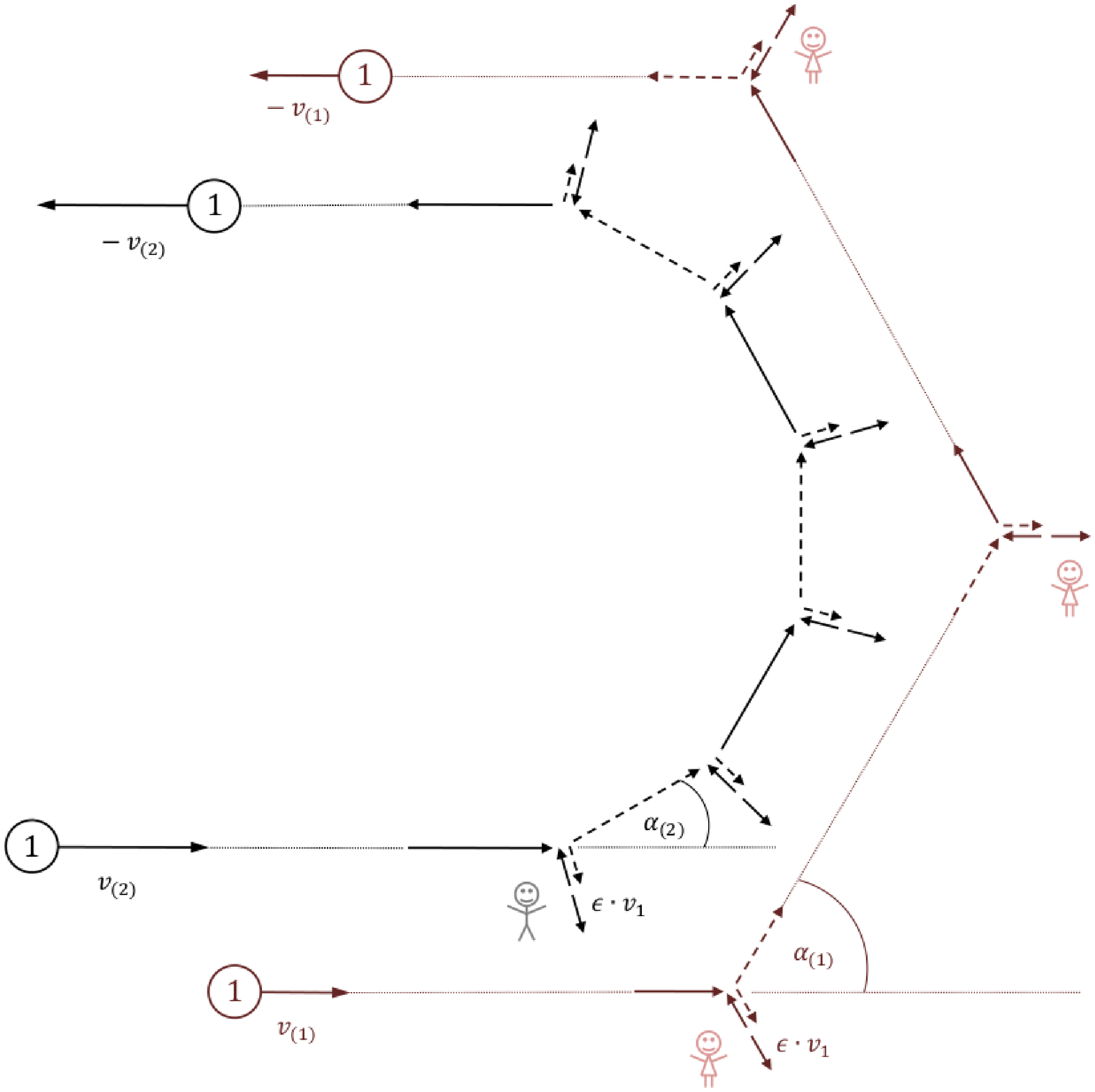}  
  \end{center}
  \vspace{-0.5cm}
  \caption{\label{pic_SRT_impulse_reversion} In a coordinated effort Alice and Bob's team of physicists steer a series of transversal standard kicks $w_{T}$ to reverse the impulse of particle $\circMunit_{\:v_{(1)}}$ resp. $\circMunit_{\:v_{(2)}}$
    }
  \end{figure}
Three assistants have to line up at the corners and initiate each steering kick timely and at suitable position. After a series of three transversal kicks
\be
   W_{(1)} \; := \;  w_{T}^{(30^{\circ})} \ast w_{T}^{(90^{\circ})} \ast w_{T}^{(150^{\circ})} \nn
\ee
against the same object $\circMunit_{\:v_{(1)}}$, which between the kicks moves freely with same velocity $v_{(1)}$, its motion gets reversed. Bob's team steers a separate reversion process for the faster particle $\circMunit_{\:v_{(2)}}$ with velocity $v_{(2)}>v_{(1)}$ which requires twice the standard kicks (\ref{Abschnitt -- kin quant elast coll - prepare congruent transversal impulse}). Six men line up at the corners and know how to fire reservoir elements $\circMunit$ into its way
\be
   W_{(2)} \; := \; w_{T}^{(15^{\circ})} \ast w_{T}^{(45^{\circ})} \ast \ldots \ast
   w_{T}^{(165^{\circ})}  \;\; . \nn
\ee
After six successive kicks of the same strength its direction of motion is reversed too.

Alice and Bob align the reversion processes $W_{(1)}$ and $W_{(2)}$ for the three incident objects. Alice rotates her reversion process for the first incident particle $\circMunit_{\:v_{(1)}}$ by $\beta=195^{\circ}$
\[
   \mathrm{R}_{\beta}\!\left[ w_{T}^{(30^{\circ})} \ast w_{T}^{(90^{\circ})} \ast w_{T}^{(150^{\circ})} \right] \; = \; w_{T}^{(30^{\circ}+\beta)} \ast w_{T}^{(90^{\circ}+\beta)} \ast w_{T}^{(150^{\circ}+\beta)} \;\; \mathrm{;}
\]
for the second incident particle $\circMunit_{\:v_{(1)}}$ she rotates the entire configuration $\mathrm{R}_{165^{\circ}}\!\left[W_{(1)}\right]$ by an angle $\beta=165^{\circ}$. Her assistants rebuild the same model from the same building blocks in a modified orientation (symbolized by operator $\mathrm{R}_{\beta}[\;\cdot\;]$). For every transversal steering kick $w_{T}^{(\theta)} \; := \;  w_{\epsilon}^{(\theta)} \ast \; w_T$ they pick two resting unit objects $\circMunit_{\:\mathbf{0}}$ from the reservoir and generate two recoil particles $\circMunit_{-\epsilon\cdot\mathbf{v}_{\mathbf{1}}}$ with same velocity $-\epsilon\cdot\mathbf{v}_{\mathbf{1}}$: one in the preparation $w_{\epsilon}^{(\theta)}$ (see figure \ref{pic_exact_annihilation1}a) and the other after the elastic kick $w_T$ (see figure \ref{pic_exact_annihilation1}b). In order to retrieve those resources Alice and Bob align their reversion processes
\be\label{Abschnitt -- kin quant elast coll - associate three impulse reversion processes}
   W_{(2)} \; \ast \; \mathrm{R}_{165^{\circ}}\!\left[W_{(1)}\right] \; \ast \; \mathrm{R}_{195^{\circ}}\!\left[W_{(1)}\right]
\ee
such that all transversal standard kicks pair up along the dashed lines in diametrically opposed locations (see figure \ref{pic_SRT_composition_coarse_grained}a)
\[
   \left(w_{T}^{(15^{\circ})} \ast w_{T}^{(30^{\circ}+165^{\circ})}\right) \ast
   \left(w_{T}^{(45^{\circ})} \ast w_{T}^{(30^{\circ}+195^{\circ})}\right) \ast \ldots \ast
   \left(w_{T}^{(165^{\circ})} \ast w_{T}^{(150^{\circ}+195^{\circ})}\right) \;\; .
\]
The four antiparallel recoil particles $\circMunit_{\:\epsilon\cdot\mathbf{v}_{\mathbf{1}}}$, $\circMunit_{\:\epsilon\cdot\mathbf{v}_{\mathbf{1}}}$, $\circMunit_{-\epsilon\cdot\mathbf{v}_{\mathbf{1}}}$ and $\circMunit_{-\epsilon\cdot\mathbf{v}_{\mathbf{1}}}$ recharge the two - temporarily expended - standard springs $\mathcal{S}_{\epsilon}\big|_{\mathbf{v}=\mathbf{0}}$ and return four resting particles back into the external reservoir $\left\{ \mathcal{S}_{\epsilon}\big|_{\mathbf{0}} , \circMunit_{\:\mathbf{0}} \right\}$ (see figure \ref{pic_exact_annihilation1}c). In the end the reservoir remains unaltered.
The net process (\ref{Abschnitt -- kin quant elast coll - associate three impulse reversion processes}) provides an elastic collision between three equivalent objects:
\be
   \circMunit_{\:\mathbf{v}_{(2)}} \,,\: \circMunit_{\:\mathrm{R}_{15^{\circ}} \mathbf{v}_{(1)}} \,,\: \circMunit_{\:\mathrm{R}_{-15^{\circ}}\mathbf{v}_{(1)}} \;\Rightarrow\; \circMunit_{-\mathbf{v}_{(2)}} \,,\: \circMunit_{-\mathrm{R}_{15^{\circ}}\mathbf{v}_{(1)}} \,,\: \circMunit_{-\mathrm{R}_{-15^{\circ}}\mathbf{v}_{(1)}}  \;\; .  \nn
\ee
In the final state their motion is exactly reversed (see figure \ref{pic_SRT_composition_coarse_grained}a). Alice and Bob mediate their elastic repulsion by well-defined resources from an external reservoir. Those were temporarily expended but finally all recycled back. Every act of their procedure is reversible.
\\

For \textbf{step II} we refine the building blocks. We model the elastic head-on collision between one unit object $\circMunit_{\:v_{(n)}}$ from left and a spreading bundle of $n$ unit objects $\circMunit_{\,\mathrm{R}_{\theta_1} v_{(1)}}, \ldots ,\circMunit_{\,\mathrm{R}_{\theta_n} v_{(1)}}$ from right. From the layout of our standard building blocks we determine the admissible velocities $v_{(n)}$ resp. $v_{(1)}$ and the suitable orientations $\theta_k$ for $k=1,\ldots,n$ (see figure \ref{pic_SRT_composition_coarse_grained}b).

Alice and Bob refine the strength $\epsilon$ of their radial standard kicks $w_{T}$ (\ref{Formel - transversal standard kick}). Each reservoir element $\circMunit_{\:\epsilon\cdot v_{\mathbf{1}}}$ deflects incident particle $\circMunit_{\:v_{(i)}}$ with admissible velocity $v_{(i)}$ by corresponding angle $\alpha_{i}$ $i=1,n$. Let Alice concatenate $N_{(1)}:=\frac{\pi}{\alpha_{1}}$ radial standard kicks
\be\label{Abschnitt -- kin quant elast coll - associate N_(1) transversal collisions}
   W_{(1)} \; := \;  w_{T}^{( -\frac{\alpha_1}{2} + \alpha_1 )} \ast w_{T}^{( -\frac{\alpha_1}{2} + 2\cdot \alpha_1 )} \ast \ldots \ast w_{T}^{( -\frac{\alpha_1}{2} + N_{(1)}\cdot\alpha_1 )}
\ee
to reverse the motion for each element $\circMunit_{\:v_{(1)}}$ of the right incident bundle. Similarly Bob steers $N_{(n)}:=\frac{\pi}{\alpha_n}$ radial kicks of \emph{same} strength $\epsilon$ like Alice
\be
   W_{(n)} \; := \; w_{T}^{( -\frac{\alpha_n}{2} + \alpha_n )} \ast w_{T}^{( -\frac{\alpha_n}{2} + 2\cdot \alpha_n )} \ast \ldots \ast w_{T}^{( -\frac{\alpha_n}{2} + N_{(n)}\cdot\alpha_n )}  \nn
\ee
until the direction of motion for the left particle $\circMunit_{\:v_{(n)}}$ with velocity $v_{(n)}$ is reversed too.

For alignment both reversion processes $W_{(1)}$ and $W_{(n)}$ must match with one another. We impose \emph{matching} deflection angles
\be\label{Abschnitt -- kin quant elast coll - matching condition}
   \alpha_1 \stackrel{!}{=} n\cdot \alpha_n \;\; .
\ee
Then for fixed radial impact velocity $\epsilon\cdot\mathbf{v}_{\mathbf{1}}$ of the reservoir element $\circMunit_{\:\epsilon\cdot\mathbf{v}_{\mathbf{1}}}$ and deflection angle $\alpha_i$ the admissible velocities $v_{(i)}$ of the incident particle $\circMunit_{\:v_{(i)}}$ $i=1,n$ are known (\ref{Formel - v-alpha-elastic transversal collision}).

Let Alice align the $n$ bundle elements $\circMunit_{\,\mathrm{R}_{\theta_1} v_{(1)}}, \ldots ,\circMunit_{\,\mathrm{R}_{\theta_n} v_{(1)}}$ from right with velocity $v_{(1)}$
\begin{itemize}
\item   under orientations $\theta_k:= \frac{n+1}{2}\cdot\alpha_n - k \cdot \alpha_n$ for $k=1,\ldots,n$ \footnote{For step I with $n=2$, $\alpha_1=60^{\circ}$, $\alpha_2=30^{\circ}$ we verify $\theta_1:=\frac{3}{2}\cdot \alpha_2 - \alpha_2 = \frac{1}{2}\cdot \alpha_2$ and $\theta_2:=\frac{3}{2}\cdot \alpha_2 - 2\cdot\alpha_2 = -\frac{1}{2}\cdot \alpha_2$ in accordance with figure \ref{pic_SRT_composition_coarse_grained}a.}
\item   with equal spacing $\Delta\theta = \alpha_n$ ranging between $\theta_1=+\frac{\alpha_1}{2}-\frac{\alpha_n}{2}$ ,  $\theta_n=-\frac{\alpha_1}{2}+\frac{\alpha_n}{2}$ .
\end{itemize}
Then Alice turns the reversion process (\ref{Abschnitt -- kin quant elast coll - associate N_(1) transversal collisions}) for first bundle element $\circMunit_{\,\mathrm{R}_{\theta_1}v_{(1)}}$
\[
   \mathrm{R}_{\beta_1}\!\left[ w_{T}^{(\vartheta_1)} \ast \ldots \ast w_{T}^{(\vartheta_{N_{(1)}})} \right] \; = \; w_{T}^{(\vartheta_1+\beta_1)} \ast \ldots \ast w_{T}^{(\vartheta_{N_{(1)}}+\beta_1)}
\]
with $\vartheta_j:= -\frac{\alpha_1}{2}+j\cdot \alpha_1$ for $j=1,\ldots,N_{(1)}$ around angle $\beta_1:=\pi+\underbrace{\frac{\alpha_1}{2}-\frac{\alpha_n}{2}}_{=:\theta_1}$ and similarly she turns the reversion process $\mathrm{R}_{\beta_k}\!\left[W_{(1)}\right]$ for every other element $\circMunit_{\,\mathrm{R}_{\theta_k}v_{(1)}}$ around angle $\beta_k:=\pi+\theta_k$ for $k=1,\ldots,n$. Similar to figure \ref{pic_SRT_composition_coarse_grained}a Alice and Bob connect reversion processes
\be\label{Abschnitt -- kin quant elast coll - associate n+1 impulse reversion processes}
   W_{(n)} \; \ast \; \mathrm{R}_{\beta_1}\!\left[W_{(1)}\right] \; \ast \; \ldots \; \ast \; \mathrm{R}_{\beta_n}\!\left[W_{(1)}\right]
\ee
such that all radial steering kicks with $\gamma_l:= -\frac{\alpha_n}{2}+l\cdot \alpha_n$ for $l=1,\ldots,N_{(n)}$
\[
\begin{array}{l}
   \left\{w_{T}^{(\gamma_1)} \ast \ldots \ast w_{T}^{(\gamma_{N_{(n)}})}\right\}
    \;\ast\; \left\{ w_{T}^{(\vartheta_1+\beta_1)} \ast \ldots \ast w_{T}^{(\vartheta_{N_{(1)}}+\beta_1)} \right\} \nn \\
   \;\;\;\;\;\;\;\;\;\;\;\;\;\;\;\;\;\;\;\;\;\;\;\;\;\;\;\;\;\;\;\;\;\;\;\;\;\;\;\;\;\;\;\; \ast \; \ldots
   \;\ast\; \left\{ w_{T}^{(\vartheta_1+\beta_n)} \ast \ldots \ast w_{T}^{(\vartheta_{N_{(1)}}+\beta_n)}  \right\} \nn
\end{array}
\]
divide into antiparallel pairs\footnote{Straight forward insertion confirms that first pair is aligned antiparallel and analogous for all the rest
\bea
   \gamma_1 - (\delta_1+\beta_n) & := &  -\frac{\alpha_n}{2}+1\cdot \alpha_n \;\; - \;\; \left( -\frac{\alpha_1}{2}+1\cdot \alpha_1 \;\; + \;\; \pi + \frac{n+1}{2}\cdot\alpha_n - n \cdot \alpha_n \right) \nn \\
   & = &  \frac{\alpha_n}{2} \; - \; \frac{\alpha_1}{2} \; - \; \pi + \frac{n\cdot\alpha_n}{2} - \frac{\alpha_n}{2} \;\; \stackrel{(\ref{Abschnitt -- kin quant elast coll - matching condition})}{=} \;\; -\pi \;\; .\nn
\eea
} where as before all byproducts can be retrieved
\[
   \left(w_{T}^{(\gamma_1)} \ast w_{T}^{(\delta_1+\beta_n)}\right) \ast
   \left(w_{T}^{(\gamma_{2})} \ast w_{T}^{(\delta_1+\beta_{n-1})}\right) \ast \ldots \ast
   \left(w_{T}^{(\gamma_{N_{(n)}})} \ast w_{T}^{(\delta_{N_{(1)}}+\beta_1)}\right) \;\; .
\]
The net process (\ref{Abschnitt -- kin quant elast coll - associate n+1 impulse reversion processes}) mediates the elastic collision of $n+1$ equivalent objects
\be
   \circMunit_{\:v_{(n)}} \,,\: \circMunit_{\:\mathrm{R}_{\theta_1} v_{(1)}} \,, \ldots ,\: \circMunit_{\:\mathrm{R}_{\theta_n} v_{(1)}}
   \;\Rightarrow\;
   \circMunit_{-v_{(n)}} \,,\: \circMunit_{-\mathrm{R}_{\theta_1} v_{(1)}} \,, \ldots ,\: \circMunit_{-\mathrm{R}_{\theta_n} v_{(1)}}  \;\; ; \nn
\ee
their motion is exactly reversed.
\\

In \textbf{step III} we refine the building blocks in model (\ref{Abschnitt -- kin quant elast coll - associate n+1 impulse reversion processes}) to the limit $\epsilon\rightarrow 0$ where the impact of individual reservoir elements $\circMunit_{\:\epsilon\cdot\mathbf{v}_{\mathbf{1}}}$ diminishes. Each radial standard kick $w_T$ (\ref{Formel - transversal standard kick}) deflects the right bundle element $\circMunit_{\:v_{\mathbf{1}}}$ with \emph{fixed} velocity $v_{(1)}\stackrel{!}{=}v_{\mathbf{1}}$ by angle $\sin \frac{\alpha_{1}}{2} \stackrel{(\ref{Formel - v-alpha-elastic transversal collision})}{:=} \frac{\epsilon}{v_{\mathbf{1}}}$ and the left particle $\circMunit_{\:v_{(n)}}$ by matching angle $\alpha_n\stackrel{(\ref{Abschnitt -- kin quant elast coll - matching condition})}{:=}\frac{1}{n}\cdot\alpha_1$. We integrate an increasing number $N_{(1)}:=\frac{\pi}{\alpha_{1}}$ resp. $N_{(n)}:=n\cdot N_{(1)}$ of refined standard kicks into the model until the motion of every particle from the right bundle and from the left is reversed. In return the spreading of the bundle $\theta_1-\theta_n := \lim_{\epsilon\rightarrow0} \; (n-1) \cdot \alpha_{1}(v_{\mathbf{1}},\epsilon) = 0$ narrows.

We rewrite the matching condition $\alpha_{1}(v_{\mathbf{1}},\epsilon) \stackrel{!}{=} n\cdot\alpha_{n}(v_{n},\epsilon)$ between the deflection angles of Alice and Bob's radial steering kicks
\be\label{Abschnitt -- kin quant elast coll - associate n+1 - matching deflection refinement}
   \lim_{\epsilon\rightarrow0} \frac{\alpha_1}{2} \; = \; \lim_{\epsilon\rightarrow0} \; \underbrace{\sin\left(\frac{\alpha_1}{2}\right)}_{\stackrel{(\ref{Formel - v-alpha-elastic transversal collision})}{=}\:\frac{\epsilon}{v_\mathbf{1}}}
   \;\; \stackrel{!}{=} \;\; \lim_{\epsilon\rightarrow0} \; n\cdot\frac{\alpha_{n}}{2}
   \;=\; n\cdot \lim_{\epsilon\rightarrow0} \; \underbrace{\sin\left(\frac{\alpha_n}{2}\right)}_{\stackrel{(\ref{Formel - v-alpha-elastic transversal collision})}{=}\:\frac{\epsilon}{v_{n}}}
\ee
and find the admissible velocity $v_{n}(v_{\mathbf{1}},\epsilon)$ for the left particle $\circMunit_{\:v_{n}}$ from $\frac{1}{v_{\mathbf{1}}} \stackrel{(\ref{Abschnitt -- kin quant elast coll - associate n+1 - matching deflection refinement})}{=} n \cdot \frac{1}{\lim_{\epsilon\rightarrow0} \; v_n}$. We approximate (\ref{Abschnitt -- kin quant elast coll - associate n+1 impulse reversion processes}) the elastic head-on collision between one standard object $\circMunit_{\:\mathbf{v}_{n}}$ and a parallel beam of $n$ elements $\{\circMunit_{\:\mathbf{v}_{\mathbf{1}}}, \ldots , \circMunit_{\:\mathbf{v}_{\mathbf{1}}} \}\equiv \circMn_{\:\mathbf{v}_{\mathbf{1}}}$ (see figure \ref{pic_SRT_composition_coarse_grained}c). Before and after the collision they fly with same velocity $\mathbf{v}_{\mathbf{1}}$ as if they were bound in a composite $\circMn_{\:\mathbf{v}_{\mathbf{1}}}$
\be
   \circMunit_{\:\mathbf{v}_n} \,,\: \circMn_{\:\mathbf{v}_{\mathbf{1}}} \;\;\stackrel{w_H}{\Rightarrow}\;\; \circMunit_{-\mathbf{v}_n} \,,\: \circMn_{-\mathbf{v}_{\mathbf{1}}} \nn \;\;.
\ee
In the limit when the bundle becomes a ray the initial velocities $\mathbf{v}_{\mathbf{1}}$,$\,\mathbf{v}_n$ satisfy relation
\be
   \lim_{\epsilon\rightarrow0} \; \mathbf{v}_n \;\; = \; - n \cdot \mathbf{v}_{\mathbf{1}} \;\; . \nn
\ee
\qed
We learn something about elastic collisions which we did not presuppose before. Our model provides a physical derivation of fundamental (collision) equation $m_1\cdot \Delta v_1 = m_2 \cdot \Delta v_2$ (including scope and limitations).

\subsection{Calorimeter absorption model}\label{Kap - KM_Dyn - Calorimeter model}

\begin{figure}    
  \begin{center}           
  \includegraphics[height=18cm]{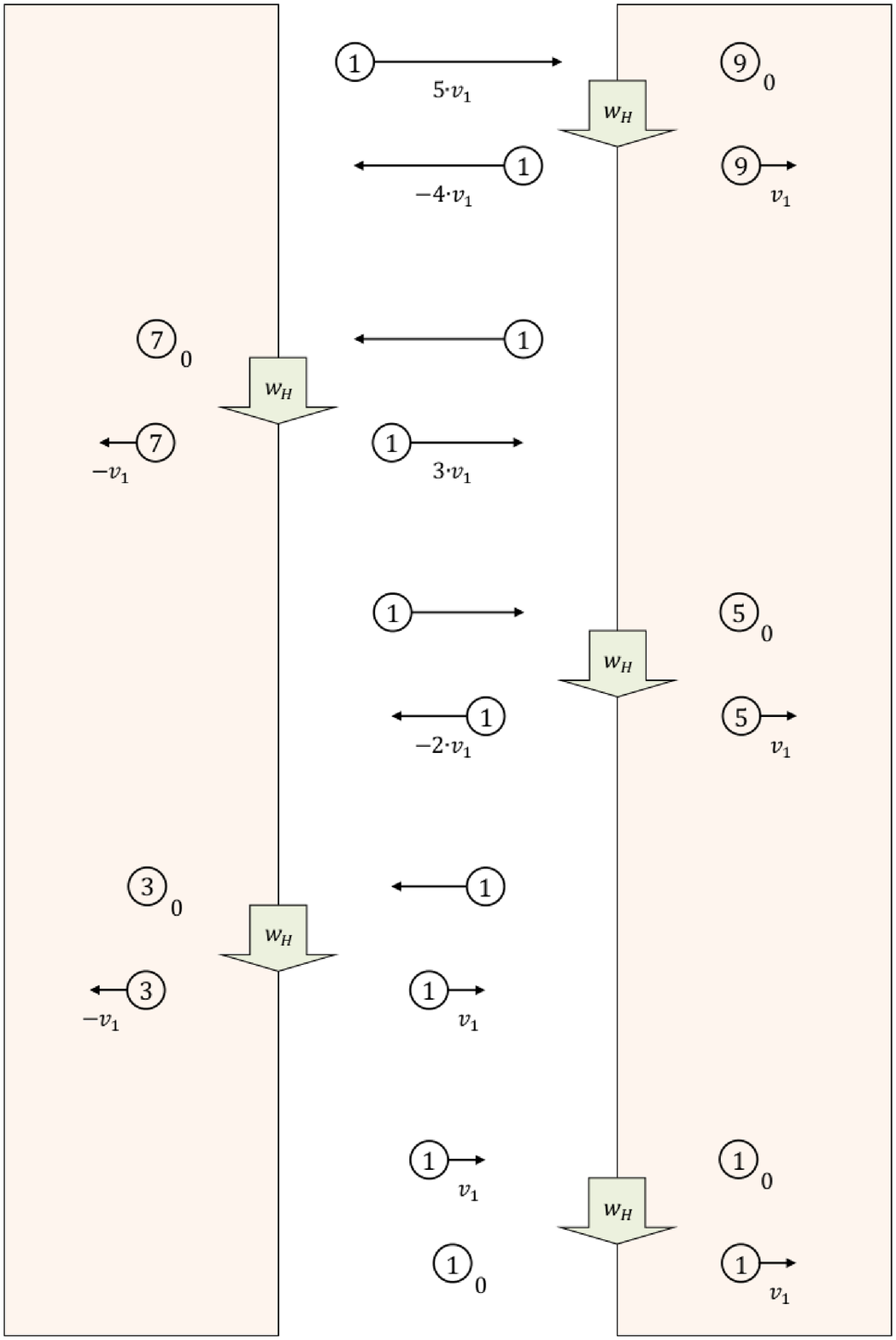}  
  \end{center}
  \vspace{-0cm}
  \caption{\label{pic_calorimeter_model} incident particle successively comes to rest by means of  elastic collisions with initially resting elements on the left resp. right side of the calorimeter reservoir $\left\{ \circMunit_{\:\mathbf{v}=\mathbf{0}} \right\}$
    }
  \end{figure}

Consider for example the elastic head-on collision $w_H$ (\ref{Abschnitt -- kin quant elast coll - elast head-on collision}) between one (fast) standard particle and a composite of $9$ elements. For a drive-by observer the incident particle kicks a resting composite into motion $\mathbf{v}$ and rebounds with reduced velocity to the left (see figure \ref{pic_calorimeter_model}). From those decelerations kicks we build our calorimeter model $W_{\mathrm{cal}}$. On the left we place again a suitable number of $7$ reservoir elements into the way, such that they get kicked out with the same standard velocity $\mathbf{v}$. The incident particle successively rebounds with reduced velocity, until (after five right- and left-deceleration kicks) it stops inside the calorimeter. For the controlled deceleration of a particle $\circMunit_{\:\textcolor{cyan}{5}\cdot \mathbf{v}}$ with velocity $5\cdot \mathbf{v}$ we mobilize a total of $25$ initially resting reservoir elements. We kick $10$ particle pairs $\left\{\circMunit_{\:\mathbf{v}}, \circMunit_{-\mathbf{v}}\right\}$ with the same standard velocity $\pm\mathbf{v}$ out of both sides of the calorimeter and $5$ single recoil particles $\circMunit_{\:\mathbf{v}}$
\be
   \circMunit_{\:\textcolor{cyan}{5}\cdot \mathbf{v}} \,,\: \textcolor{magenta}{25}\cdot \circMunit_{\:\mathbf{0}} \;\;\;
   \Rightarrow \;\;\;
   \circMunit_{\:\textcolor{cyan}{\mathbf{0}}} \,,\: \textcolor{magenta}{10} \cdot \left\{\circMunit_{\:\mathbf{v}}, \circMunit_{-\mathbf{v}}\right\} \,,\:  \textcolor{magenta}{5}\cdot \circMunit_{\:\mathbf{v}} \nn \;\; .
\ee
We formulate this process as ''reaction equation'' (along the language use among chemists).\footnote{Physicists formulate equations between \emph{measures} (\emph{Ma\ss{}}gleichungen); chemists on the contrary transitions between their \emph{carriers} (Ma\ss{}\emph{tr\"ager}) - we formulate both sides: carrier and its measure!} If we absorb the same standard particle $\circMunit_{\:\textcolor{cyan}{8}\cdot \mathbf{v}}$ with higher velocity $8\cdot \mathbf{v}$, we have to mobilize even $64$ initially resting reservoir elements. In the same series of deceleration kicks
\be
   \circMunit_{\:\textcolor{cyan}{8}\cdot \mathbf{v}} \,,\: \textcolor{magenta}{64}\cdot \circMunit_{\:\mathbf{0}} \;\;\;
   \Rightarrow \;\;\;
   \circMunit_{\:\textcolor{cyan}{\mathbf{0}}} \,,\: \textcolor{magenta}{28} \cdot \left\{\circMunit_{\:\mathbf{v}}, \circMunit_{-\mathbf{v}}\right\} \,,\:  \textcolor{magenta}{8}\cdot \circMunit_{\:\mathbf{v}}  \nn
\ee
we generate $28$ standard particle pairs and $8$ impulse carriers.

This illustrates the \emph{essence} of a basic measurement. The vague pre-theoretic comparison $\circMunit_{\:\textcolor{cyan}{8}\cdot \mathbf{v}} >_{E,\mathbf{p}} \circMunit_{\:\textcolor{cyan}{5}\cdot \mathbf{v}}$ ''more absorption effect than'' and ''overrunning in head-on collision'' is now precisely determined by a \emph{number of equivalent reference elements} ($\textcolor{magenta}{28}$ resp. $\textcolor{magenta}{10}$ particle pairs $\left\{\circMunit_{\:\mathbf{v}}, \circMunit_{-\mathbf{v}}\right\}$ of equal capability to work and $\textcolor{magenta}{8}$ resp. $\textcolor{magenta}{5}$ recoil particles $\circMunit_{\:\mathbf{v}}$ of equal impact). We will assess the physical meaning of the extracted calorimeter elements as units of energy and momentum by pre-theoretic ordering relations in Proposition \ref{Prop - kin quant Absorptions Wirkung - pre-theoretical characterization E unit and p unit}.
\begin{pr}
The calorimeter-deceleration-cascade $W_{\mathrm{cal}}$ is a physical model for absorbing unit object $\circMunit_{\:\textcolor{cyan}{n}\cdot \mathbf{v}_{\mathbf{1}}}$ with velocity $n\cdot \mathbf{v}_{\mathbf{1}}$ in an external calorimeter where it comes to rest $\circMunit_{\:\mathbf{0}}$
\be
   \circMunit_{\:\textcolor{cyan}{n}\cdot \mathbf{v}_{\mathbf{1}}}  \;\;
   \stackrel{W_{\mathrm{cal}}}{\Rightarrow} \;\;
   \circMunit_{\:\textcolor{cyan}{\mathbf{0}}} \,,\:
   \mathrm{RB} \nn \;\; .
\ee
In return we extract the reservoir balance for absorption
\be\label{Abschnitt -- kin quant Absorptions Wirkung - Reservoirbilanz - absorption}
   \mathrm{RB}\left[\circMunit_{\:n\cdot \mathbf{v}_{\mathbf{1}}} \Rightarrow \circMunit_{\:\mathbf{0}} \right] \;\; := \;\;
   \left( \frac{1}{2}\cdot n^2 - \frac{1}{2}\cdot n  \right)\cdot \left\{\circMunit_{-\mathbf{v}_{\mathbf{1}}}, \circMunit_{\:\mathbf{v}_{\mathbf{1}}}\right\} \;\; , \;\; n\cdot \circMunit_{\:\mathbf{v}_{\mathbf{1}}}
\ee
a certain number of standard particle pairs $\left\{\circMunit_{-\mathbf{v}_{\mathbf{1}}}, \circMunit_{\:\mathbf{v}_{\mathbf{1}}}\right\}$ and impulse carriers $\circMunit_{\:\mathbf{v}_{\mathbf{1}}}$ from a reservoir with resting standard elements $\left\{ \circMunit_{\:\mathbf{0}} \right\}$ (which we suppress in the notation).
\end{pr}
\textbf{Proof:}
Let the initial velocity be $(2 i+1)\cdot \mathbf{v}_{\mathbf{1}}$. Alice steers a cascade of elastic collisions with suitable packs of resting reservoir elements. We need to adjust the deceleration kicks.

Let Bob pick a suitable head-on collision $w_H$, which satisfies the relation $\mathbf{v} \stackrel{(\ref{Abschnitt -- kin quant elast coll - kinemtical relations elast collision two generic objects})}{=} - n\cdot \mathbf{w}$ (with initial velocities $\mathbf{v}:=(2i+\frac{1}{2})\cdot \mathbf{v}_{\mathbf{1}}$, $\mathbf{w}:=-\frac{1}{2}\cdot \mathbf{v}_{\mathbf{1}}$ and $n:=4i+1$). We pick the numerical values for later convenience. Thus Bob prepares a composite $\underbrace{\circMunit_{\:\mathbf{0}}\ast \ldots \ast \circMunit_{\:\mathbf{0}}}_{(4i+1)\times}$ of $(4i+1)$ elements such that in an elastic head-on collision (\ref{Abschnitt -- kin quant elast coll - elast head-on collision})
\be\label{Abschnitt -- kin quant Absorptions Wirkung - elastic longitudinal collision - antisym}
   \circMunit_{\:(2i+\frac{1}{2})\cdot \mathbf{v}_{\mathbf{1}}} \,,\: (4i+1)\cdot\circMunit_{-\frac{1}{2}\cdot\mathbf{v}_{\mathbf{1}}} \;\;\stackrel{w_H}{\Rightarrow}\;\; \circMunit_{-(2i+\frac{1}{2})\cdot \mathbf{v}_{\mathbf{1}}} \,,\: (4i+1)\cdot\circMunit_{\:\frac{1}{2}\cdot\mathbf{v}_{\mathbf{1}}}
\ee
the incident particle $\circMunit_{\:(2i+\frac{1}{2})\cdot \mathbf{v}_{\mathbf{1}}}$ with velocity $(2i+\frac{1}{2})\cdot \mathbf{v}_{\mathbf{1}}$ rebounds antiparallel from the composite $(4i+1)\cdot\circMunit_{-\frac{1}{2}\cdot\mathbf{v}_{\mathbf{1}}}$ with velocity $-\frac{1}{2}\cdot\mathbf{v}_{\mathbf{1}}$.

Let $\mathcal{B}$ob move relative to $\mathcal{A}$lice with constant velocity $\mathbf{v}_{\mathcal{B}}=\frac{1}{2}\cdot\mathbf{v}_{\mathbf{1}^{(\mathcal{A})}}$ to the right. She will see $\mathcal{B}$ob's collision with the same number of colliding elements, but different measured values of velocity (Galilei covariant transformation $v_{\circMunit}^{(\mathcal{A})}= v_{\circMunit}^{(\mathcal{B})} + v_{\mathcal{B}}^{(\mathcal{A})}$ for every element $\circMunit$). For $\mathcal{A}$lice the incident particle $\circMunit_{\:(2i+1)\cdot \mathbf{v}_{\mathbf{1}}}$ kicks into the right side of the calorimeter with velocity $(2i+1)\cdot \mathbf{v}_{\mathbf{1}}$ and rebounds with reduced velocity $-2i\cdot \mathbf{v}_{\mathbf{1}}$ to the left
\be\label{Abschnitt -- kin quant Absorptions Wirkung - elastic longitudinal collision - rechts}
   w_{H,r} : \;\; \circMunit_{\:(2i+1)\cdot \mathbf{v}_{\mathbf{1}}} \,,\: (4i+1)\cdot\circMunit_{\:\mathbf{0}} \;\;\stackrel{(\ref{Abschnitt -- kin quant Absorptions Wirkung - elastic longitudinal collision - antisym})}{\Rightarrow}\;\; \circMunit_{-2i\cdot \mathbf{v}_{\mathbf{1}}} \,,\: (4i+1)\cdot\circMunit_{\:\mathbf{v}_{\mathbf{1}}}
\ee
while the resting composite $(4i+1)\cdot\circMunit_{\:\mathbf{0}}$ gets kicked into standard velocity $\mathbf{v}_{\mathbf{1}}$. On the left side Alice places a new composite of $(4j+1)$ elements and generates the next analog deceleration kick
\be
   w_{H,l} : \;\; \circMunit_{-(2j+1)\cdot \mathbf{v}_{\mathbf{1}}} \,,\: (4j+1)\cdot\circMunit_{\:\mathbf{0}} \;\;\stackrel{(\ref{Abschnitt -- kin quant Absorptions Wirkung - elastic longitudinal collision - rechts})}{\Rightarrow}\;\; \circMunit_{\:2j\cdot \mathbf{v}_{\mathbf{1}}} \,,\: (4j+1)\cdot\circMunit_{-\mathbf{v}_{\mathbf{1}}} \;\; . \nn
\ee
with $j:=i-\frac{1}{2}$. After each round of right and left collisions $W := w_{H,r} \ast w_{H,l}$
\be\label{Abschnitt -- kin quant Absorptions Wirkung - elastic longitudinal collision - Umkehrabfolge}
   \circMunit_{\:(2i+1)\cdot \mathbf{v}_{\mathbf{1}}} \,,\: (4i+1)\cdot\circMunit_{\:\mathbf{0}} \,,\: (4i-1)\cdot\circMunit_{\:\mathbf{0}}
   \;\;\stackrel{W}{\Rightarrow}\;\;
   \circMunit_{\:(2i-1)\cdot \mathbf{v}_{\mathbf{1}}} \:,\;  (4i+1)\cdot\circMunit_{\:\mathbf{v}_{\mathbf{1}}} \,,\: (4i-1)\cdot\circMunit_{-\mathbf{v}_{\mathbf{1}}}
\ee
we add the extracted reservoir elements $(4i-1)\cdot\circMunit_{-\mathbf{v}_{\mathbf{1}}}\,,\: (4i+1)\cdot\circMunit_{\:\mathbf{v}_{\mathbf{1}}}$ from both sides of the calorimeter and the successive deceleration $\Delta v \stackrel{(\ref{Abschnitt -- kin quant Absorptions Wirkung - elastic longitudinal collision - Umkehrabfolge})}{:=} -2\cdot v_{\mathbf{1}}$. Each antiparallel particle pair $\circMunit_{-\mathbf{v}_{\mathbf{1}}} ,  \circMunit_{\:\mathbf{v}_{\mathbf{1}}} \stackrel{w_{\mathbf{1}}^{-1}}{\Rightarrow} \mathcal{S}_{\mathbf{1}}\big|_{\mathbf{0}} , \circMunit_{\:\mathbf{0}} , \circMunit_{\:\mathbf{0}}$ can charge a spring $\mathcal{S}_{\mathbf{1}}\big|_{\mathbf{0}}$ by standard process $w_{\mathbf{1}}^{-1}$  (\ref{Abschnitt -- basic dynamical measures - Einheitswirkung}). The resting elements stay in the calorimeter reservoir $\left\{ \circMunit_{\:\mathbf{0}} \right\}$. On each deceleration step $W_{(i)}$ $i=1,\dots,N$ Alice extracts
\be\label{Abschnitt -- kin quant Absorptions Wirkung - Reservoirbilanz - Deceleration}
   \mathrm{RB}\left[\circMunit_{\:(2i+1)\cdot \mathbf{v}_{\mathbf{1}}} \Rightarrow \circMunit_{\:(2i-1)\cdot \mathbf{v}_{\mathbf{1}}} \right] \; \stackrel{(\ref{Abschnitt -- kin quant Absorptions Wirkung - elastic longitudinal collision - Umkehrabfolge}) }{:=} \;
   (4i-1)\cdot  \mathcal{S}_{\mathbf{1}}\big|_{\mathbf{0}} \;\; , \;\; 2 \cdot \circMunit_{\:\mathbf{v}_{\mathbf{1}}}
\ee
$(4i-1)\cdot  \mathcal{S}_{\mathbf{1}}\big|_{\mathbf{0}}$ standard springs and $2\cdot \circMunit_{\:\mathbf{v}_{\mathbf{1}}}$ single elements (see figure \ref{pic_calorimeter_model}).\footnote{Two consecutive deceleration rounds and a final clean-up kick bring incident particle $\circMunit_{\:5\cdot \mathbf{v}_{\mathbf{1}}}$ with velocity $5\cdot \mathbf{v}_{\mathbf{1}}$ to rest. Alice counts $7+3$ particle pairs $\left\{\circMunit_{\:\mathbf{v}}, \circMunit_{-\mathbf{v}}\right\}$ and $2+2+1$ impulse carriers $\circMunit_{\:\mathbf{v}_{\mathbf{1}}}$.}

For the initial velocity $(2N+1)\cdot \mathbf{v}_{\mathbf{1}}$ Alice steers $N$ consecutive deceleration rounds $W_{\mathrm{cal}} \; := \; W_{(1)} \ast\dots\ast W_{(N)}$ inside the calorimeter until the incident object $\circMunit_{\:(2N+1)\cdot \mathbf{v}_{\mathbf{1}}}$ stops. Alice extracts the total reservoir balance for absorption  (\ref{Abschnitt -- kin quant Absorptions Wirkung - Reservoirbilanz - absorption})
\bea
   \mathrm{RB}\left[\circMunit_{\:(2N+1)\cdot \mathbf{v}_{\mathbf{1}}} \Rightarrow \circMunit_{\: \mathbf{0}} \right] & = &
   \sum_{i=1}^{N} \mathrm{RB}\left[\circMunit_{\:(2i+1)\cdot \mathbf{v}_{\mathbf{1}}} \Rightarrow \circMunit_{\:(2i-1)\cdot \mathbf{v}_{\mathbf{1}}} \right]
   \; + \; \mathrm{RB}\left[\circMunit_{\:\mathbf{v}_{\mathbf{1}}} \Rightarrow \circMunit_{\: \mathbf{0}} \right] \nn \\
   & \stackrel{(\ref{Abschnitt -- kin quant Absorptions Wirkung - Reservoirbilanz - Deceleration})}{=} & \sum_{i=1}^{N} \left( (4i-1) \cdot \mathcal{S}_{\mathbf{1}}\big|_{\mathbf{0}} \: , \; 2 \cdot \circMunit_{\:\mathbf{v}_{\mathbf{1}}} \right)
   \; + \; \left( 0\cdot \mathcal{S}_{\mathbf{1}}\big|_{\mathbf{0}} \: , \; 1 \cdot \circMunit_{\:\mathbf{v}_{\mathbf{1}}} \right) \nn \\
   & = & \!\!\! \underbrace{(2\cdot N^2+N)}_{\frac{1}{2}\cdot (2N + 1)^2 - \frac{1}{2}\cdot (2N +1)} \!\!\!\!
   \cdot \; \mathcal{S}_{\mathbf{1}}\big|_{\mathbf{0}} \;\: , \;\; (2N+1) \cdot \circMunit_{\:\mathbf{v}_{\mathbf{1}}}  \;\; .  \nn
\eea
\qed

\section{Quantification}\label{Kap - KM_Dyn - Quantification}

Now let us consider these models from an abstract physical perspective. We define the basic observables from elemental ordering relations \{\ref{Kap - KM_Dyn - Basic observable}\}. They fix the \emph{physical meaning} of our reference objects as units for energy and momentum and of our calorimeter extract.
\begin{pr}\label{Prop - kin quant Absorptions Wirkung - pre-theoretical characterization E unit and p unit}
Our standard spring $\mathcal{S}_{\mathbf{1}}\big|_{\mathbf{v}=\mathbf{0}}$ represents the unit energy and has no impulse.
\be\label{Abschnitt -- kin quant Absorptions Wirkung - pre-theoretical characterization E unit and p unit}
\begin{array}{rclrcl}
   E \left[ \mathcal{S}_{\mathbf{1}}\big|_{\mathbf{0}} \right] & &
   & \;\;\;\;\;\;\;\;\;\;\;\;\;
   E \left[ \circMunit_{\:\mathbf{v}_\mathbf{1}} \right] & = & \frac{1}{2}\cdot E \left[ \mathcal{S}_{\mathbf{1}}\big|_{\mathbf{0}} \right]
   \\
   \mathbf{p} \left[ \mathcal{S}_{\mathbf{1}}\big|_{\mathbf{0}} \right] & = & 0
   & \mathbf{p} \left[ \circMunit_{\:\mathbf{v}_\mathbf{1}} \right] &  &
\end{array}
\ee
The standard impulse carrier $\circMunit_{\:\mathbf{v}_\mathbf{1}}$ represents the unit momentum and also has energy.
\end{pr}
\textbf{Proof:}
The two dimensions energy and impulse are inseparably intertwined in unit action $\mathcal{S}_{\mathbf{1}}\big|_{\mathbf{0}} , \circMunit_{\:\mathbf{0}} , \circMunit_{\:\mathbf{0}} \stackrel{w_{\mathbf{1}}}{\Rightarrow} \circMunit_{-\mathbf{v}_{\mathbf{1}}} , \circMunit_{\:\mathbf{v}_{\mathbf{1}}}$ between our standard energy source and impulse carriers.

The resting energy source $\mathcal{S}_{\mathbf{1}}\big|_{\mathbf{v}=\mathbf{0}}$ can not overrun any moving object in a head-on collision test; it has no impact. Its abstract momentum vanishes $\mathbf{p} \left[ \mathcal{S}_{\mathbf{1}}\big|_{\mathbf{0}} \right] = 0$ (see Definition \ref{Def - vortheor Ordnungsrelastion - impulse}).

We can convert two comoving elements $\left\{\circMunit_{\:\mathbf{v}_{\mathbf{1}}}, \circMunit_{\:\mathbf{v}_{\mathbf{1}}}\right\} \Rightarrow \left\{\circMunit_{-\mathbf{v}_{\mathbf{1}}}, \circMunit_{\:\mathbf{v}_{\mathbf{1}}}\right\}$ into an antiparallel particle pair by letting, vividly spoken, one element repulse elastically $\circMunit_{\:\mathbf{v}_{\mathbf{1}}} \,,\: \circMM_{\:\mathbf{v}_M} \stackrel{(\ref{Abschnitt -- kin quant elast coll - elast head-on collision})}{\Rightarrow} \circMunit_{-\mathbf{v}_{\mathbf{1}}} \,,\: \circMM_{-\mathbf{v}_M}$ from a much heavier ''reservoir block''. In the limit $m[\circMunit] \ll m[\circMM]$ one can show by refined calorimeter measurements that the ''bouncing block'' $\mathbf{v}_M\rightarrow 0$ is practically at rest and with negligible contribution to energy (for transitivity and details see \cite{Hartmann-diss}). Thus two impulse units generate the same \emph{absorption effect} like one standard spring; by the equipollence of cause and effect (Definition \ref{Def - vortheor Ordnungsrelastion - energie}) their energy is the same $2\cdot E \left[ \circMunit_{\:\mathbf{v}_{\mathbf{1}}} \right] = E \left[ \mathcal{S}_{\mathbf{1}}\big|_{\mathbf{0}} \right]$.
\qed

The calorimeter extract has the same capability to execute work as the incident particle, because our calorimeter model is reversible.\footnote{We steer the absorption by a series of elastic head-on collisions in system $\circMunit_{\:n\cdot \mathbf{v}_{\mathbf{1}}} \cup \left\{ \circMunit_{\:\mathbf{0}} \right\}$ of incident particle and calorimeter reservoir. Every step of the deceleration cascade is reversible, because it is build up from solely congruent standard actions $w_{\mathbf{1}}$ and because physicists can steer these processes both ways.} It also has the same impact, since otherwise one could construct a Perpetuum Mobile.
\begin{lem}\label{Lem - kin quant Absorptions Wirkung - Reservoirbilanz - p conserved}
In our calorimeter model $W_{\mathrm{cal}}$ the extracted impulse carriers $\circMunit\ast\ldots\ast\circMunit_{\;\mathbf{v}_{\mathbf{1}}}$
\be
   \circMunit\ast\ldots\ast\circMunit_{\;\mathbf{v}_{\mathbf{1}}} \;\; \sim_{\mathbf{p}} \;\; \circMa_{\:\mathbf{v}_{a}}   \nn
\ee
have same impulse as incident particle $\circMa_{\:\mathbf{v}_{a}}$. The transferred momentum is \underline{conserved}.
\end{lem}
\textbf{Proof:} Without restricting generality we consider the absorption (\ref{Abschnitt -- kin quant Absorptions Wirkung - Reservoirbilanz - absorption}) of one standard particle $\circMa\equiv\circMunit$ with velocity $\mathbf{v}_{a}:=-n\cdot \mathbf{v}_{\mathbf{1}}$, which extracts $n$ standard impulse carriers $\circMunit\ast\ldots\ast\circMunit_{\;\mathbf{v}_{\mathbf{1}}} =: \circMn_{\:\mathbf{v}_{\mathbf{1}}}$ with velocity $\mathbf{v}_{\mathbf{1}}$. In a generic inelastic head-on collision test the incident particle $\circMa_{\:\mathbf{v}_{a}}$ and its impulse model $\circMn_{\:\mathbf{v}_{\mathbf{1}}}$
\be\label{Abschnitt -- Impuls - inelastic collision direct}
   \circMa_{\:\mathbf{v}_{a}} \,,\: \circMn_{\:\mathbf{v}_{\mathbf{1}}} \;\; \stackrel{w_{(d)}}{\Rightarrow} \;\; \circMa\ast\circMn_{\:\mathbf{v}'} \;\; .
\ee
form a bound aggregate $\circMa\ast\circMn_{\:\mathbf{v}'}$ with velocity $\mathbf{v}'$. They have the same momentum if they collide, stick together and come to rest (see \emph{physical} Definition \ref{Def - vortheor Ordnungsrelastion - impulse}). Hence, we have to show, that the bound aggregate $\circMa\ast\circMn_{\:\mathbf{v}'}$ must stop $\mathbf{v}'\stackrel{!}{=}\mathbf{0}$.

Let us hypothetically assume it moves. Then we could stop the aggregate $\circMa\ast\circMn_{\:\mathbf{v}'}$ in our calorimeter
\be
   \circMa\ast\circMn_{\;\mathbf{v}'} \;\; \stackrel{W_{\mathrm{cal}}}{\Rightarrow} \;\; \circMa\ast\circMn_{\;\mathbf{0}} \,,\: k\cdot \mathcal{S}_{\mathbf{1}}\big|_{\mathbf{0}} \,,\: l\cdot \circMunit_{\:\mathbf{v}_{\mathbf{1}}}    \nn
\ee
and extract additional $k$ energy units $\mathcal{S}_{\mathbf{1}}\big|_{\mathbf{0}}$ and $l$ impulse carriers $\circMunit_{\:\mathbf{v}_{\mathbf{1}}}$ into the direction of $\mathbf{v}'$. But then we could set up a circular process (from reversible standard actions) which could kick impulse carriers $\circMunit_{\:\mathbf{v}_{\mathbf{1}}}$ ''around the corner'' $\circMunit_{\:\mathrm{R}_{\theta}\mathbf{v}_{\mathbf{1}}}$ without effecting anything else (details see \cite{Hartmann-diss}).

This hypothetical process violates Euler's principle of sufficient reason; that every change of motion requires an external cause (physical reason) \cite{Euler Anleitung} and impossibility of a perpetuum mobile. A moving observer (with velocity $\mathbf{v}_{\mathbf{1}}$) could set initially resting reservoir particles $\{\circMunit_{\:\mathbf{0}}\}$ into motion with velocity $2\cdot\mathbf{v}_{\mathbf{1}}$ but also into opposite direction with velocity $-2\cdot\mathbf{v}_{\mathbf{1}}$ and thus generate particle pairs $\{\circMunit_{-2\cdot\mathbf{v}_{\mathbf{1}}} \,,\: \circMunit_{\:2\cdot\mathbf{v}_{\mathbf{1}}} \} \sim_{E} \mathcal{S}_{2}\big|_{\mathbf{0}}$ resp. energy sources without any reaction from nothing. Hence $\mathbf{v}' \stackrel{!}{=} 0$ must have been zero; in the inelastic head-on collision test $w_{(d)}$ the particle $\circMa_{\:\mathbf{v}_{a}}$ and its impulse model $\circMn_{\:\mathbf{v}_{\mathbf{1}}}$ stick together and come to rest.
\qed

The kinetic energy of the incident particle $E\left[\circMa_{\:\mathbf{v}_a}\right]$ is completely \emph{transformed} into potential energy of the absorber material; the latter comes in congruent portions $\mathcal{S}_{\mathbf{1}}\big|_{\mathbf{0}}$. According to the \emph{congruence principle}
\be \nn
   E\left[ \circMa_{\:\mathbf{v}_a} \right] \;\; \stackrel{(\mathrm{Def.}\ref{Def - vortheor Ordnungsrelastion - energie})}{=} \;\;  E\left[ \,\mathcal{S}_{\mathbf{1}}\big|_{\mathbf{0}} \ast \ldots \ast \mathcal{S}_{\mathbf{1}}\big|_{\mathbf{0}} \, \right]
   \; \stackrel{(\mathrm{Congr.})}{=:} \; E \cdot E\left[ \mathcal{S}_{\mathbf{1}}\big|_{\mathbf{0}} \right]
\ee
we measure ''how many times larger'' its kinetic energy is, than the potential energy of one standard spring $\mathcal{S}_{\mathbf{1}}\big|_{\mathbf{0}}$. The numerical factor $E := \sharp \left\{ \mathcal{S}_{\mathbf{1}}\big|_{\mathbf{0}} \right\}$ is a \emph{physical quantity}; our reference energy $E\left[ \mathcal{S}_{\mathbf{1}}\big|_{\mathbf{0}} \right]$ is a \emph{dimension}.

In the same way we conduct \emph{independent} basic measurements of momentum and inertial mass. Sommerfeld \cite{Sommerfeld-Mechanik} defines the impulse of a moving body: ''Impulse means (with regards to direction and magnitude) that kick, which is capable of generating a given state of motion from the initial state of rest.''
Our absorption model (\ref{Abschnitt -- kin quant Absorptions Wirkung - Reservoirbilanz - absorption}) provides a direct measurement. We generate this kick by a number $\mathbf{p}:=\sharp \left\{ \circMunit_{\:\mathbf{v}_{\mathbf{1}}} \right\}$ of congruent standard kicks\footnote{Sommerfeld's defining kick is associated with generating motion (from rest). We examine kicks which annihilate motion (towards rest). In two-body collisions Sommerfeld regards the ''recipient''; we the ''giver''.}
\be \nn
   \mathbf{p}\left[ \circMa_{\:\mathbf{v}_a} \right] \;\; \stackrel{(\mathrm{Def.}\ref{Def - vortheor Ordnungsrelastion - impulse})}{=} \;\;  \mathbf{p}\left[ \, \circMunit_{\;\mathbf{v}_{\mathbf{1}}}\!\ast\ldots\ast\circMunit_{\;\mathbf{v}_{\mathbf{1}}} \, \right]
   \; \stackrel{(\mathrm{Congr.})}{=:} \; \mathbf{p} \cdot \mathbf{p}\left[ \circMunit_{\;\mathbf{v}_{\mathbf{1}}} \right]  \;\; .
\ee
Similarly we measure the inertia of body $\circMa$ with an equally massive (Definition \ref{Def - vortheor Ordnungsrelastion - inertial mass}) composite of standard elements and the latter according to the congruence principle
\be
   m\left[ \circMa \right] \;\;\stackrel{(\mathrm{Def.}\ref{Def - vortheor Ordnungsrelastion - inertial mass})}{=}\;\;
   m\left[ \circMunit \ast \dots \ast \circMunit \right]  \;\;\stackrel{(\mathrm{Congr.})}{=:}\;\;
   m\cdot m\left[ \circMunit \right]   \nn
\ee
by the number $m:= \sharp \left\{ \circMunit \right\} $ of standard elements and their unit mass $m\left[ \circMunit \right]$.

\section{Quantity equations}\label{Kap - KM_Dyn - Quantity equations}

In the calorimeter model we can count the individual standard springs, impulse carriers, reservoir elements etc. We derive the relation between these physical quantities; the primary dynamical equations.
\begin{lem}\label{Lem - kin quant Absorptions Wirkung - Reservoirbilanz - absorption proportional materiemenge}
Let a composite of $m$ bound unit elements $\circMunit \ast \dots \ast \circMunit\;\sim_{m^{\mathrm{(inert)}}}\circMa$ have the same inertia as a generic particle $\circMa_{\:\mathbf{v}_{a}}$. Then the reservoir extract for absorbing the particle
\be\label{Abschnitt -- kin quant Absorptions Wirkung - Reservoirbilanz - absorption proportional materiemenge}
   \mathrm{RB} \left[ \circMa_{\:\mathbf{v}_{a}} \Rightarrow
   \circMa_{\:\mathbf{0}}  \right] \;\; = \;\; m \cdot \mathrm{RB} \left[ \circMunit_{\:\mathbf{v}_{a}} \Rightarrow \circMunit_{\:\mathbf{0}}  \right]
\ee
is $m$ times larger than for absorbing one unit element $\circMunit_{\:\mathbf{v}_{a}}$ with the same velocity $\mathbf{v}_{a}$.
\end{lem}
\textbf{Proof:}
Same inertia (Definition \ref{Def - vortheor Ordnungsrelastion - inertial mass}) implies, that in an elastic head-on collision test with same initial velocity $v'_{a}:=\frac{1}{2}\cdot v_{a}$ the composite $\circMunit\ast\dots\ast\circMunit$ and the generic particle $\circMa$ must repulse in the same anti-symmetrical way $w_H: \circMa_{\:\mathbf{v}'_{a}} \,,\; \circMunit\ast\dots\ast\circMunit_{\:-\mathbf{v}'_{a}} \Rightarrow \circMa_{-\mathbf{v}'_{a}} \,,\; \circMunit\ast\dots\ast\circMunit_{\;\mathbf{v}'_{a}}$. For an observer moving with relative velocity $-\mathbf{v}'_{a}$ \be
   \circMa_{\:\mathbf{v}_{a}} \,,\; \circMunit\ast\dots\ast\circMunit_{\;\mathbf{0}} \;\; \stackrel{w_H}{\Rightarrow} \;\; \circMa_{\:\mathbf{0}} \,,\; \circMunit\ast\dots\ast\circMunit_{\;\mathbf{v}_{a}} \nn
\ee
the generic particle $\circMa_{\:\mathbf{v}_{a}}$ stops and kicks the initially resting composite into same motion.

We neutralize this elastic head-on collision by absorbing the (spectator) composite in our calorimeter
$\mathrm{RB} \left[ \circMunit\ast\dots\ast\circMunit_{\:\mathbf{v}_{a}} \Rightarrow \circMunit\ast\dots\ast\circMunit_{\:\mathbf{0}} \right] =: k_m\cdot \mathcal{S}_{\mathbf{1}}\big|_{\mathbf{0}} \,,\, l_m\cdot\,\circMunit_{\:\mathbf{v}_{\mathbf{1}}}$ and by catapulting the (temporarily) resting particle in a reversed absorption $-\mathrm{RB} \left[ \circMa_{\:\mathbf{v}_{a}} \Rightarrow \circMa_{\:\mathbf{0}} \right] =: k_a\cdot \mathcal{S}_{\mathbf{1}}\big|_{\mathbf{0}} \,,\, l_a\cdot\,\circMunit_{\:\mathbf{v}_{\mathbf{1}}}$ back into the original motion. The net effect is a circular process. The net reservoir extract
\be\label{Abschnitt -- kin quant Absorptions Wirkung - Reservoirbilanz - absorption proportional materiemenge - hilfssatz}
   \mathbf{p} \left[ (k_a-k_m)\cdot\mathcal{S}_{\mathbf{1}}\big|_{\mathbf{0}} \:,\; (l_a-l_m)\cdot\circMunit_{\:\mathbf{v}_{\mathbf{1}}} \right] \;\; \stackrel{(\ref{Abschnitt -- kin quant Absorptions Wirkung - pre-theoretical characterization E unit and p unit})}{=} \;\;  \mathbf{p} \left[ (l_a-l_m)\cdot\circMunit_{\:\mathbf{v}_{\mathbf{1}}} \right] \;\; \stackrel{!}{=} \;\; 0
\ee
cannot have momentum (for conservation see Lemma \ref{Lem - kin quant Absorptions Wirkung - Reservoirbilanz - p conserved}). It also cannot have energy
\be
   E \;[ \:(k_a-k_m)\cdot\mathcal{S}_{\mathbf{1}}\big|_{\mathbf{0}} \:,\; \underbrace{(l_a-l_m)\cdot\circMunit_{\:\mathbf{v}_{\mathbf{1}}}}_{\stackrel{(\ref{Abschnitt -- kin quant Absorptions Wirkung - Reservoirbilanz - absorption proportional materiemenge - hilfssatz})}{=}\; 0} \:] \;\; \stackrel{!}{=} \;\; 0  \nn \;\; .
\ee
Thus the generic particle $\circMa_{\:\mathbf{v}_{a}}$ generates the same reservoir extract ($l_a \stackrel{!}{=} l_m$ impulse carriers
$\circMunit_{\:\mathbf{v}_{\mathbf{1}}}$ and $k_a \stackrel{!}{=} k_m$ energy units $\mathcal{S}_{\mathbf{1}}\big|_{\mathbf{0}}$) as for absorbing the composite
\be
   \mathrm{RB} \left[ \circMa_{\:\mathbf{v}_{a}} \Rightarrow \circMa_{\:\mathbf{0}} \right] \;\;=\;\;
   \mathrm{RB} \left[ \circMunit\ast\dots\ast\circMunit_{\;\mathbf{v}_{a}} \Rightarrow \; \circMunit\ast\dots\ast\circMunit_{\;\mathbf{0}} \right] \;\;=\;\;
   m \cdot \mathrm{RB} \left[ \circMunit_{\:\mathbf{v}_{a}} \Rightarrow \circMunit_{\:\mathbf{0}} \right] \nn \;\; .
\ee
From the pack we absorb every element one by one. All extracted energy-momentum units are congruent; their total number adds up to $m$ times the output of one unit element $\circMunit_{\:\mathbf{v}_{a}}$.
\qed
\begin{theo}\label{Theorem - kin quant absorption action for free particle}
The particle $\circMa_{\:\mathbf{v}_{a}}$ with inertial mass $m_{\circMa}=m\cdot m_{\circMunit}$ and velocity $\mathbf{v}_a=n\cdot \mathbf{v}_{\mathbf{1}}$ has a \underline{kinetic} energy and momentum
\bea\label{Abschnitt -- kin quant Absorptions Wirkung - kin energy and momentum metrisiert}
   E \left[ \circMa_{\:\mathbf{v}_{a}} \right] & = & \left( \frac{1}{2} \cdot m \cdot n^2   \right) \;\cdot\; E \left[ \mathcal{S}_{\mathbf{1}}\big|_{\mathbf{0}} \right]
   \\
   \mathbf{p} \left[ \circMa_{\:\mathbf{v}_{a}} \right] & = & \left( m \cdot n \right)\; \cdot\; \mathbf{p} \left[ \circMunit_{\:\mathbf{v}_{\mathbf{1}}} \right]   \;\; . \nn
\eea
\end{theo}
\textbf{Proof:}
By reversibility and equipollence principle the calorimeter absorption extract has the same ''capability to work'' as the incident particle $\circMa_{\:\mathbf{v}_{a}}$. Its kinetic energy is transformed
\bea
   E \left[ \circMa_{\:\mathbf{v}_{a}} \right] & \stackrel{(\mathrm{Equip.})}{=} & E \left[ \mathrm{RB} \left[ \circMa_{\:\mathbf{v}_{a}} \Rightarrow \circMa_{\:\mathbf{0}}  \right] \right] \nn \\
   & \stackrel{(\ref{Abschnitt -- kin quant Absorptions Wirkung - Reservoirbilanz - absorption proportional materiemenge})}{=} & E \left[  m \cdot \mathrm{RB} \left[ \circMunit_{\:\mathbf{v}_{a}} \Rightarrow \circMunit_{\:\mathbf{0}} \right] \right] \nn \\
   & \stackrel{(\ref{Abschnitt -- kin quant Absorptions Wirkung - Reservoirbilanz - absorption})(\mathrm{Congr.})}{=} &
   m \cdot \left\{ \left(\frac{1}{2} \cdot n^2 - \frac{1}{2}\cdot n \right)\cdot E \left[ \mathcal{S}_{\mathbf{1} }\big|_{\mathbf{0}} \right] \;\; + \;\; n\cdot E \left[ \circMunit_{\:\mathbf{v}_{\mathbf{1}}} \right] \right\} \nn \\
   & \stackrel{(\ref{Abschnitt -- kin quant Absorptions Wirkung - pre-theoretical characterization E unit and p unit})}{=} &  \left( \frac{1}{2} \cdot m \cdot n^2   \right) \;\cdot\; E \left[ \mathcal{S}_{\mathbf{1} }\big|_{\mathbf{0}} \right]   \nn
\eea
into the potential energy of $( \frac{1}{2} \cdot m \cdot n^2 )$ congruent energy units $\mathcal{S}_{\mathbf{1}}\big|_{\mathbf{0}}$. The calorimeter extract has the same impulse as the incident particle $\circMa_{\:\mathbf{v}_{a}}$ (see Lemma \ref{Lem - kin quant Absorptions Wirkung - Reservoirbilanz - p conserved}). Its impulse
\bea
   \mathbf{p} \left[ \circMa_{\:\mathbf{v}_{a}} \right] & = & \mathbf{p} \left[ \mathrm{RB} \left[ \circMa_{\:\mathbf{v}_{a}} \Rightarrow \circMa_{\:\mathbf{0}}  \right] \right] \nn \\
   & \stackrel{(\ref{Abschnitt -- kin quant Absorptions Wirkung - Reservoirbilanz - absorption})(\ref{Abschnitt -- kin quant Absorptions Wirkung - pre-theoretical characterization E unit and p unit})}{=} & \mathbf{p} \left[ m \cdot n \cdot \circMunit_{\:\mathbf{v}_{\mathbf{1}}} \right] \;\; \stackrel{(\mathrm{Congr.})}{=} \;\; \left( m \cdot n \right) \;\cdot\; \mathbf{p} \left[ \circMunit_{\:\mathbf{v}_{\mathbf{1}}} \right]  \nn
\eea
is reproduced by $( m\cdot n )$ congruent impulse units $\circMunit_{\:\mathbf{v}_{\mathbf{1}}}$ from the calorimeter reservoir.
\qed
When we build the model in Galilei kinematics we derive primary dynamical equations
(\ref{Abschnitt -- kin quant Absorptions Wirkung - kin energy and momentum metrisiert})
\[
   \frac{E_{a}}{E_{\mathbf{1}}} = \frac{1}{2} \cdot \frac{m_a}{m_{\mathbf{1}}} \cdot \left(\frac{\mathbf{v}_a}{ \mathbf{v}_{\mathbf{1}}}\right)^2 \;\;\;\;\;\;\;\;
   \frac{\mathbf{p}_a}{\mathbf{p}_{\mathbf{1}}} = \frac{m_a}{m_{\mathbf{1}}} \cdot \frac{\mathbf{v}_a}{ \mathbf{v}_{\mathbf{1}}} \;\;,
\]
in which all numerical values for energy $E =: \frac{E \left[ \circMa_{\:\mathbf{v}_{a}} \right]}{E \left[ \mathcal{S}_{\mathbf{1}}\big|_{\mathbf{0}} \right] }\:$, impulse $\mathbf{p} =: \frac{\mathbf{p} \left[ \circMa_{\:\mathbf{v}_{a}} \right]}{\mathbf{p} \left[ \circMunit_{\:\mathbf{v}_{\mathbf{1}}} \right]}\:$, mass $m =: \frac{m\left[ \circMa \right]}{m \left[ \circMunit \,\right]}\:$ and velocity $n =: \frac{v_a}{v_{\mathbf{1}}}$ occur in the form \emph{measure/unit measure}. Each formal ratio symbolizes the result of a physical operation; counting congruent units in the calorimeter model. When we steer the \emph{same} measurement process in Poincare kinematics, then we will derive all equations of relativistic dynamics \cite{Hartmann-SRT_Dyn}.

We measure the momentum from multi-partite systems by steering a separate absorption $W^{(i)}_{\mathrm{cal}}$ for each element. We extract impulse carriers $\circMunit_{\:\mathbf{v}_{\mathbf{1}}}$ and $\circMunit_{-\mathbf{v}_{\mathbf{1}}}$ - on the left and right side of the deceleration cascade - in the direction of its motion $\mathbf{v}_i$ (see figure \ref{pic_calorimeter_model}). For a generic many-particle system one extracts impulse carriers $\left\{\circMunit_{\:\mathbf{v}_{\!i}}\right\}_{i=1\ldots N}$ in various directions $\mathbf{v}_{i}\neq\mathbf{v}_{j}$.
\begin{theo}
Direction and magnitude of total momentum is calculable by \underline{vectorial addition}
\be\label{Abschnitt -- vectorial momentum addition}
    \mathbf{p}\left[ \circMunit_{\:\mathbf{v}_1} , \dots , \circMunit_{\:\mathbf{v}_N} \right]  \;\; = \;\; \mathbf{p}\left[ \circMunit_{\:\mathbf{v}_1} \right] + \ldots + \; \mathbf{p}\left[ \circMunit_{\:\mathbf{v}_N} \right]   \;\; .
\ee
\end{theo}
\textbf{Proof:}
(In Galilei kinematics) we construct a physical model $W$ for absorbing multiple standard elements $\circMunit_{\:\textcolor{cyan}{\mathbf{v}_i}}$ with velocities $\mathbf{v}_i$ into various directions $\mathbf{v}_i \nparallel \mathbf{v}_j$
\be
   \left\{\circMunit_{\:\textcolor{cyan}{\mathbf{v}_1}} , \dots , \circMunit_{\:\textcolor{cyan}{\mathbf{v}_N}}\right\} \,,\: \circMunit_{\:\mathbf{0}}
   \;\; \stackrel{W}{\Rightarrow} \;\; \left\{\circMunit_{\:\textcolor{cyan}{\mathbf{0}}} , \dots , \circMunit_{\:\textcolor{cyan}{\mathbf{0}}}\right\}  \,,\:
\circMunit_{\:\mathbf{v}_{(N)}} \;\; .
\nn
\ee
All elements $i=1,\ldots,N$ of the system $\left\{\circMunit_{\:\mathbf{0}} , \dots , \circMunit_{\:\mathbf{0}}\right\}$ stop; while one initially resting absorber particle $\circMunit_{\:\mathbf{v}_{(N)}}$ gets kicked, as we will show, into velocity $\mathbf{v}_{(N)} = \mathbf{v}_1 + \ldots + \mathbf{v}_N$.

\begin{figure}    
  \begin{center}           
  \includegraphics[height=7.3cm]{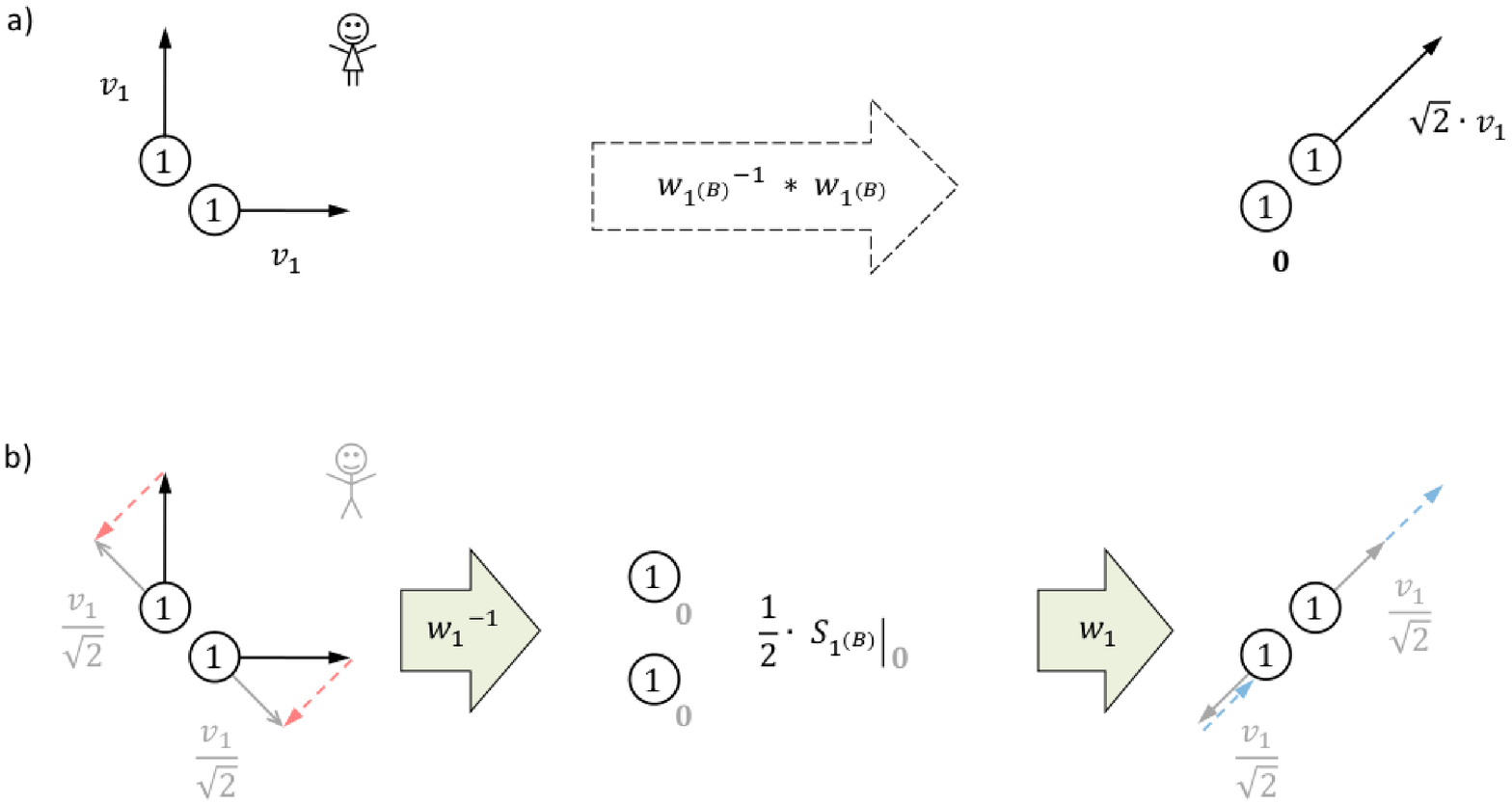}  
  \end{center}
  \vspace{-0cm}
  \caption{\label{pic_impuls_vektoriell_prinzip} a) vectorial impulse addition by underlying b) isotropic unit actions
    }
  \end{figure}

We illustrate Alice complete momentum transfer from one moving particle $\circMunit_{\:\mathbf{v}_1}$ onto another particle $\circMunit_{\:\mathbf{v}_2}$ moving into perpendicular direction in figure \ref{pic_impuls_vektoriell_prinzip}
and similarly for a system of $N$ inequivalent impulse carriers $\left\{\circMunit_{\:\mathbf{v}_1} , \circMunit_{\:\mathbf{v}_2} , \dots , \circMunit_{\:\mathbf{v}_N}\right\}$ from relativity principle and $\mathcal{B}$ob's reversible standard actions $w_{\mathbf{1}^{(\mathcal{B})}}$ by induction (details see \cite{Hartmann-diss}). At each step of his calorimeter mediated process momentum is conserved by Lemma \ref{Lem - kin quant Absorptions Wirkung - Reservoirbilanz - p conserved}. We measure the total momentum of the system $\mathbf{p}\left[ \circMunit_{\:\mathbf{v}_{1}} , \dots , \circMunit_{\:\mathbf{v}_N} \right] = \mathbf{p}\left[ \circMunit_{\:\mathbf{v}_{(N)}} \right]$ by one absorber particle and the latter in a calorimeter by the number of \emph{equivalent} impulse carriers $\circMunit_{\:\mathbf{v}_{\mathbf{1}}}$ which now all point into the \emph{same direction} of $\mathbf{v}_{(N)}:=\mathbf{v}_{1}+ \ldots + \mathbf{v}_N$. By linearity of the momentum-velocity relation (\ref{Abschnitt -- kin quant Absorptions Wirkung - kin energy and momentum metrisiert}) for a single particle the total impulse $\mathbf{p}\left[ \circMunit_{\:\mathbf{v}_1}  \dots , \circMunit_{\:\mathbf{v}_N} \right] = \left( m_{\mathbf{1}} \cdot \mathbf{v}_1 + \ldots + m_{\mathbf{1}} \cdot \mathbf{v}_N \right) \cdot \mathbf{p}[\circMunit_{\:\mathbf{v}_1}]$ is the vector sum over all elements $\left\{\circMunit_{\:\mathbf{v}_{\!i}}\right\}_{i=1\ldots N}$.
\qed

\section{Origin}\label{Kap - KM_Dyn - Origin}

Commonly one ascribes astronomical observations the central role for the development of modern natural science. Following Lorenzen \cite{Lorenzen - Entstehung der exakten Wissenschaften} we agree insofar as ''From the phenomenon of regular movement of celestial bodies humans conceived the idea of exact regularity in nature. (Though) despite precise astronomical observations, Babylonians made throughout the centuries (and despite their arithmetic rules for projections), one can not accredit to them the thought of a 'natural law' as we understand it today.'' It was a first step to the \emph{de-deification} (Entg\"otterung) of nature. We locate our initial assumptions in the historic-genetic development of everyday practical work.\footnote{According to the guiding principle of the historic school one can understand conceptual schemata of past epoches, if one recognizes ''how historic problems originate from practical problems'' in everyday life \cite{Lorenzen - Entstehung der exakten Wissenschaften}. During industrial revolution steam engine and work machine are coupled together for the first time. That marks the transition from (personal interplay of) worker and work machine to (coupling of) engine drive and work machine. While traditional craftsmen learned to handle their tools (hammer, saw, needle etc.) intuitively, industrial revolution substitutes the former by an impersonal motor. Steering the latter became an unprecedented problem. Lorenzen \cite{Lorenzen - Entstehung der exakten Wissenschaften} defines what we understand by physics today: ''is designing of mathematical theories for interpretation and prognosis of natural or \emph{technically effected processes}''. The latter is crucial for modern physics. ''Greek enlightenment did not unlike contemporary enlightenment let the \emph{ideal} - of technical mastery of nature - \emph{come into effect}.'' In this way one can understand the shift from the tradition of purely kinematical descriptions (of \emph{celestial} phenomena) to physical explanations in the proper sense (based on unprecedented mechanical concepts, inertia, force etc. from \emph{earthly} work experience).}

Already in the discussion on the principle of inertia \cite{Hartmann-SRT-Kin} we acknowledge, that Newton could draw on the literature of practical mechanics on problems of machine construction and work economy \cite{Wolff - Geschichte der Impetustheorie}. In this context we grasp an inertial reference system as a reproducible experimental prerequisite which one can isolate from the external disturbances empirically. In a billiard collision or other practical experiments the processes between the balls or the parts of a machine exceed the external e.g. gravitational or electromagnetic effects by far. In practice one can separate the internal interactions. We regard mechanics as a science of constructing and steering (local) machines. For the foundation one primarily focusses on technically \emph{controllable} natural processes.\footnote{In the ''vis viva'' dispute Leibniz \cite{Hecht - Ruben-Festschrift} was searching in ever new thought experiments on machine models (where accelerations occur) for the basic observable ''living force'' (kinetic energy). ''Starting from considerations which are concerned only with the nearest practical interests of technical work'' (\emph{extractable mechanical work} from dynamical machines, collisions, heat machines and \emph{work equivalent} from electrical, chemical or phase transitions etc.) Helmholtz \cite{Helmholtz - Ueber die Erhaltung der Kraft} was lead to discover the principle of conservation of energy - in practical form the impossibility of perpetuum mobile. With focus on work Hertz \cite{Hertz - Einleitung zur Mechanik} specified the purpose of physics: ''It is the nearest and in a sense main function of our conscious natural knowledge, that it enables us to foresee future experiences so as to be able to set up our present actions according to that foresight.''}

In this domain we define the basic observables from practical ordering relations \{\ref{Kap - KM_Dyn - Basic observable}\}. In a basic measurement \cite{Helmholtz - Zaehlen und Messen} which is always a pair comparison between the measurement object and a material model of concatenated units the latter have a different function \cite{Peter '69 - Dissertation}. While one measures the ''capability to work'' $>_E$ or ''impact'' $>_p$ of a generic interaction one must provide the \emph{reference devices} sufficiently constant \cite{Janich Das Mass der Dinge}. We use standard springs $\mathcal{S}_{\mathbf{1}}\big|_{\mathbf{0}}$ and bodies $\circMunit$ as sufficiently constant representatives of ''capability to work'' and ''impact'' \{\ref{Kap - KM_Dyn - Reference standards}\}. To derive relativistic kinematics Einstein introduces rods and clocks as unstructured entities by the clock postulate (without already having a theory of matter). In his Nobel prize lecture he ''once again tried to justify his provisional use of rods and clocks to give the geometrical statements of the theory empirical content''; though viewed it later as ''a logical shortcoming of the theory of relativity in its present form to be forced to introduce measuring rods and clocks separately instead of being able to construct them as solutions to differential equations'' \cite{Giovanelli '14 - Einstein on rods and clocks}. Our calorimeter is another example for the measurement instruments as a \emph{primitive entity} to the theory. One can substitute the provisory reference devices by refined ones, more suitable for reproducible operations. The manufacturing norms for testing the sufficient constancy of length units, energy standards etc. remain the same \cite{Hartmann-SRT-Kin}.

One can couple local calorimeters into e.g. gravitational or electromagnetic systems. Provided also a swarm of light clocks one can measure the relation between geodesic deviation and energy-momentum sources intrinsically. When in an elementary particle collision new particle generations develop, one can capture a jet with the calorimeter and measure the energy and momentum of separate decay products. We presuppose all interactions solely as a completed process with unknown inner structure. We pick the compression of a standard spring as an elementary building block for our calorimeter model because they are congruent and reproducible. Physicists can couple these units in a controlled way and count; and then measure the other interactions with this reference interaction $w_{\mathbf{1}}$.

Let us measure an electromagnetic process. After a small configuration transition $\Delta\mathbf{x}_I$ our calorimeter is sensitive to the kinetic effect $\Delta\mathbf{v}_i$ for all elements $i\in I$ of the system. We measure the associated conversion of potential energy $V_{\mathrm{pot}} \left[ \Delta\mathbf{x}_I \right] := - \sum_{i\in I} E_{\mathrm{kin}} \left[ \Delta\mathbf{v}_i \right]$ into kinetic energy of the elements. From the basic physical quantities of length, duration and energy, momentum one can define ''force'' as a derived physical quantity and ultimately derive \emph{Newton's equations of motion} for a conservative system (for details see \cite{Hartmann-diss}).

For measuring the length or duration we cover the measurement object $\mathcal{O} \sim_{t,s} \mathcal{L}\ast\ldots\ast\mathcal{L}$ by a regular grid of light clocks, which are all congruent with one another \cite{Hartmann-SRT-Kin}. For a basic measurement of interactions we build a calorimeter from one elementary collision process $w_{\mathbf{1}}$ and symmetry principles (isotropy of $w_{\mathbf{1}}$, relativity). Building the collision models is a simple steering task. We couple the building blocks $w_{\mathbf{1}}\ast\ldots\ast w_{\mathbf{1}} \sim_{E,p} w_{EM}$ to generate the same kinetic effects (change of motion $\Delta \mathbf{v}_i$) like from the electromagnetic interaction $w_{EM}$. The calorimeter model matches the generic interaction (neither more nor less energy-momentum gain ''$\sim_{E,p}$''). Despite different inner structure, with regards to the associated ''capability to work'' and ''impact'' the two processes are indistinguishable. We build the models from congruent energy sources $\sharp \left\{\mathcal{S}_{\mathbf{1}}\right\}$ and momentum carriers $\sharp \left\{\circMunit_{\:\mathbf{v}_{\mathbf{1}}}\right\}$; by counting them we find ''how many times'' more energy the generic process $w_{EM} \left[ \Delta\mathbf{x}_I \right]$ transforms than one standard spring $\mathcal{S}_{\mathbf{1}}$ in reference process $w_{\mathbf{1}}$. We measure the magnitude of the basic observables energy and momentum (defined from elementary comparisons \{\ref{Kap - KM_Dyn - Basic observable}\}). By reversibility of the calorimeter model and impossibility of perpetuum mobile (in a circular process) the physical quantities from counting the energy and momentum units are \emph{universal}. We associate the definition of basic observables ''$>_{E,p}$'' and the quantification in the calorimeter model with the symmetry of the building blocks in a vivid way.

We construct the calorimeter model \{\ref{Kap - KM_Dyn - Calorimeter model}\} from pre-theoretic building blocks \{\ref{Kap - KM_Dyn - Reference standards}\} which are subject to \emph{physical} principles (causality, inertia, relativity, impossibility of perpetuum mobile, superposition) and \emph{methodical} principles for the constructor (elementary comparison, congruence, equipollence) as well as a \emph{social} condition: Physicists must cooperate to create the material models. They have to know, \emph{when}, \emph{where} and \emph{how} to couple the reservoir elements $\circMunit$ into the deceleration process, to generate absorption. Team work is crucial for the conduct of basic measurements. Basic physical quantities are a joint product and not generated individually. In all domains of experiment and measurement, where underlying physical and methodical principles are valid, we derive primary dynamical equations (\ref{Abschnitt -- kin quant elast coll - elast head-on collision}),(\ref{Abschnitt -- kin quant Absorptions Wirkung - kin energy and momentum metrisiert}),(\ref{Abschnitt -- vectorial momentum addition}) for the basic observables \{\ref{Kap - KM_Dyn - Basic observable}\}; they are fundamental for the mathematical framework.
\\

We have demonstrated the foundation of physics as an empirical science (in contrast to pure mathematics). This approach is diametrically opposed to an axiomatic ansatz, insofar that our starting point lies in the definitions. From them one can prove the equations of motion and even the conservation laws, while in an axiomatic approach one postulates the latter (or corresponding symmetry laws) and ultimately derives the physical observables - though as a pure operand without empirical meaning. We begin from definitions which have a practical dimension. This operationalization builds on Helmholtz fundamental analysis of measurements along with congruence principle, equipollence principle etc. which are also being construed measurement theoretically. In this sense we are dealing with a foundation of physics, but not with an explanation from mathematically formulated principles, nor from the empirically given - we explain neither from what one can say, nor from what one sees, but from what one does in order to measure. In this approach, which explains the mathematical formalism from the operationalization of basic quantities, one can address and understand scope and limitations of the formalism, with significance also for other formalisms in physics.
\\
\\
\textbf{Acknowledgements} Thank you to Bruno Hartmann sen. and Peter Ruben for introducing the research problem and essential suggestions. I also want to thank Thomas Thiemann for stimulating discussions and support and Oliver Schlaudt for clarification. This work was made possible initially by the German National Merit Foundation and finally with support by the Perimeter Institute.

\end{document}